\documentclass[aps,prr,twocolumn,superscriptaddress,floatfix,notitlepage]{revtex4-2}

\usepackage[T1]{fontenc}
\usepackage{graphicx}
\usepackage{bbm}
\usepackage{float} 
\usepackage{color}
\usepackage[usenames,dvipsnames]{xcolor}
\usepackage{amsmath} 
\usepackage{amstext}
\usepackage{latexsym}
\usepackage[colorlinks=true,citecolor=NavyBlue,linkcolor=RubineRed,urlcolor=NavyBlue]{hyperref}
\usepackage{lipsum}
\usepackage{enumitem}
\usepackage{bm}
\usepackage{amssymb}

\usepackage{mathtools}
\usepackage{subfigure}
\usepackage{mathrsfs}
\usepackage{multirow}

\definecolor{purple1}{rgb}{128,0,128}

\newcommand{\bea}{\begin{eqnarray}}
\newcommand{\ea}{\end{eqnarray}}
\newcommand{\ord}{{\cal O}}

\usepackage{orcidlink}      
\usepackage{bm}             

\begin{document}


\title{Electrodynamics of vortices in quasi-two-dimensional scalar Bose-Einstein condensates
}
\author{Seong-Ho Shinn  \orcidlink{0000-0002-2041-5292}}
\email{seongho.shin@uni.lu}
\affiliation{%
  Department of Physics and Materials Science, University of Luxembourg, L-1511 Luxembourg, Luxembourg}%

\author{Adolfo del Campo  \orcidlink{0000-0003-2219-2851}}
\email{adolfo.delcampo@uni.lu}
\affiliation{Department  of  Physics  and  Materials  Science,  University  of  Luxembourg,  L-1511  Luxembourg, Luxembourg}
\affiliation{Donostia International Physics Center,  E-20018 San Sebasti\'an, Spain}

\date{\today}

\begin{abstract} In two spatial dimensions, vortex-vortex interactions approximately vary with the logarithm of the inter-vortex distance, making it possible to describe an ensemble of vortices as a Coulomb gas. 
We introduce a duality between vortices in a quasi-two-dimensional (quasi-2D) scalar Bose-Einstein condensates (BEC) and effective Maxwell's electrodynamics.
Specifically, we address the general scenario of inhomogeneous, time-dependent BEC number density with dissipation or rotation. Starting from the Gross-Pitaevskii equation (GPE), which describes the mean-field dynamics of a quasi-2D scalar BEC without dissipation, 
we show how to map vortices in a quasi-2D scalar BEC to 2D electrodynamics beyond the point-vortex approximation, even when dissipation is present or in a rotating system.   
The physical meaning of this duality is discussed. 
\end{abstract}

\maketitle

\section{Introduction}

Topological defects are ubiquitous in physics. A symmetry-breaking second-order phase transition generally leads to the formation of topological defects that can be classified according to the topology of the vacuum manifold using homotopy groups \cite{Kibble1976,KIBBLE1980,Zurek1985,zurek1993cosmic,delCampo2014,Mermin1979,Manton2004,Kawaguchi2012}. This classification identifies different kinds of defects such as kinks, vortices, domain walls, skyrmions, etc.
Among them, $U(1)$ vortices describe pointlike singularities of a complex scalar field with quantized circulation \cite{nelson2002defects}.
Their occurrence in Bose-Einstein condensates (BEC) indicates their superfluid character \cite{Fetter01,pethick}. In such context, they can be created spontaneously by driving the transition from a normal fluid to a superfluid (e.g., by a thermal quench), as theoretically analyzed \cite{Zeng2023,thudiyangal2024} and experimentally demonstrated 
\cite{Weiler08,Chomaz2015,Goo2021,Goo2022}. An alternative mechanism to create vortices involves rotating 
the normal cloud \cite{Haljan2001} or 
the superfluid cloud, pumping angular momentum in the system \cite{Abo01,Fetter09}. Vortices in degenerate ultracold gases can also be produced by phase imprinting \cite{Matthews99,Leanhardt02,Brachmann11}, by merging independent BEC \cite{Scherer07}, by stirring laser beams \cite{Madison00,Raman01,Damski2002,Henn09,Gauthier2019}, or by making superfluid flow pass an obstacle \cite{Neely10}. 
Progress in manipulating and controlling ultracold gases makes it possible to design arbitrary patterns of vortices in a BEC sample \cite{Neely2024}.

Vortex-vortex interactions are known to scale logarithmically with the inter-vortex distance within some approximations. 
In the limit of a homogeneous BEC number density, excluding the core region of each vortex, an ensemble of static vortices in a nonrotating quasi-two-dimensional (quasi-2D) scalar BEC without dissipation can be mathematically regarded as a 2D Coulomb gas \cite{nelson2002defects}. 

For nonrotating three-dimensional (3D) ${}^4 \textrm{He}$ superfluid without dissipation, 
a duality between vortices in a thin cylindrical system and 2D electrodynamics has been reported in the limit where the superfluid density is constant outside vortices \cite{halperin1979,Ambegaokar1980}. 
Such a duality has been extended to the general 3D case by using $\phi^4$ theory 
and minimizing the action with respect to the fluctuation of the superfluid density \cite{Kleinert1989,Kleinert2008}. 
In the absence of an external potential,  the asymptotic vortex dynamics in 2D nonlinear Schr\"odinger equation has been studied \cite{NEU1990_NLSE}, and a duality between 2D electrodynamics and vortices in the (2+1) dimensional nonlinear wave equation has been put forward \cite{NEU1990_NLWE}. 
For the nonrotating scalar BEC without dissipation, 
when the BEC number density is approximately constant, it has been shown that the motion of the vortex can be described according to the nonrelativistic dynamics of strings in 3D system \cite{FISCHER1999}, 
and 
the effective Maxwell's equations in a quasi-2D system \cite{Simula2020}.  This connection is valid for an inhomogeneous time-independent BEC number density in the case of a nonrotating quasi-2D scalar BEC \cite{johansen2024}.

However, the BEC number density is zero at the core of a vortex, and thus, a vortex has a finite core size, which is about the order of the healing length (coherence length) $\xi_h \coloneqq \hbar / \sqrt{2 M g n_m}$, where $n_m$ is the mean BEC number density \cite{ginzburg1958}. 
As a result, the fluctuation of the BEC number density cannot be neglected, especially around the core of the vortex. 
To simplify the problem, the point-vortex model (PVM) has been widely used. Yet, such description cannot account for the dynamic vortex creation and annihilation processes if one uses the mean-field Gross-Pitaevskii equation (GPE) under nonrotating quasi-2D scalar BEC without dissipation \cite{Pedro2011,Richaud2021,Richaud2023,corso2024}. 
It thus remains to be established whether the connection between vortices and effective Maxwell's electrodynamics is valid in a quasi-2D scalar BEC with dissipation, beyond the PVM or under rotation.

\begin{figure*}[htbp]
    \centering
    \includegraphics[width=0.7\linewidth]{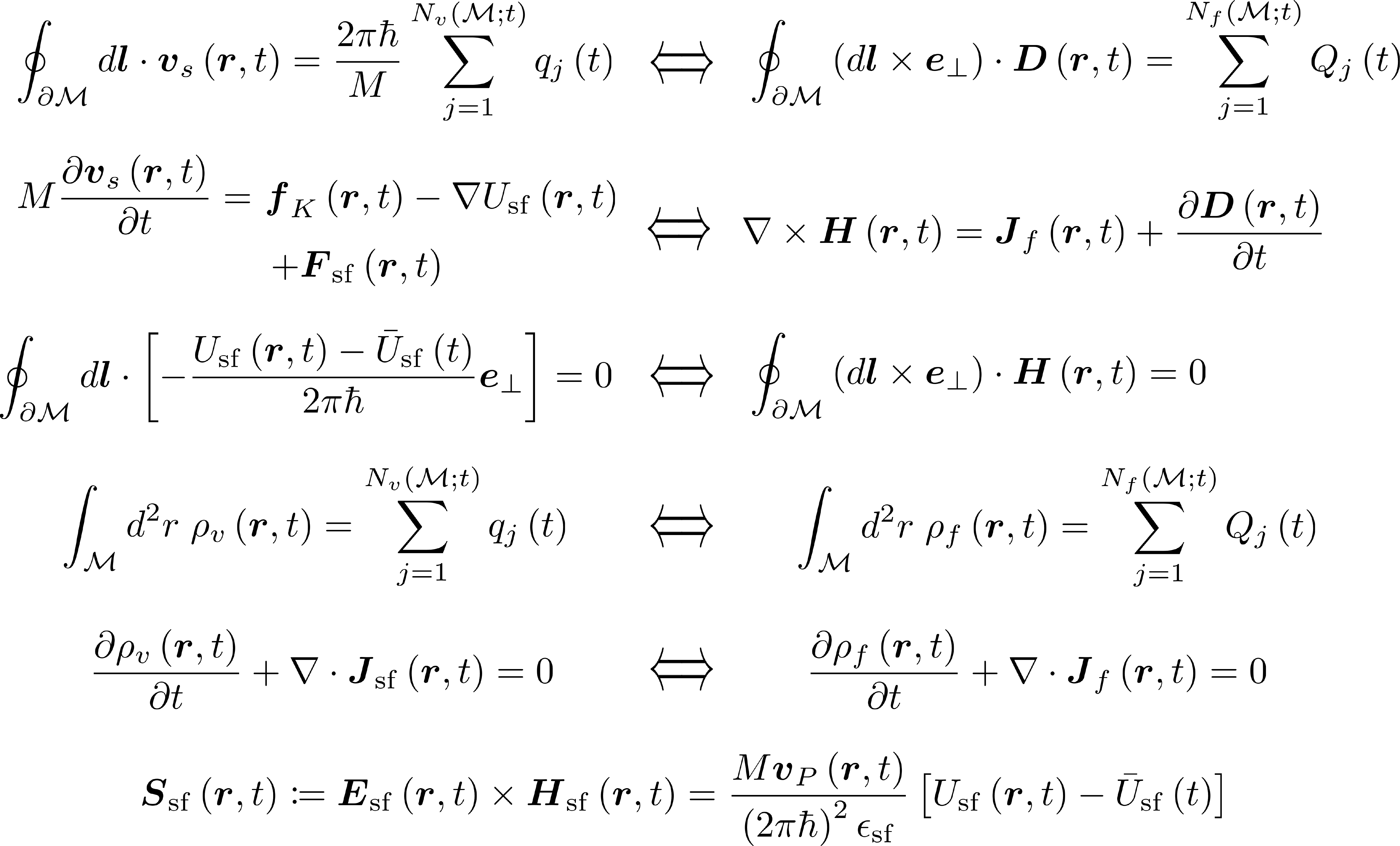}
    \caption{Duality between the vortices in the 
    quasi-2D scalar BEC and the electrodynamics in the matter. 
    The derivations are in Sec. \ref{VortexBEC_electrodynamics}, and 
    the full effective Maxwell's equations are shown in Table \ref{table:vortex-Maxwell}. 
    On the right-hand side, $\boldsymbol{D} \left( \boldsymbol{r}, t \right)$ is the electric displacement field, $\boldsymbol{H} \left( \boldsymbol{r}, t \right)$ is the magnetic field strength, $Q_j \left( t \right)$ is the free electric charge with index $j$ at time $t$, and $\boldsymbol{J}_f \left( \boldsymbol{r}, t \right)$ is the free electric current density. For other definitions of symbols, refer to Tables \ref{SuppSymbolDef} and \ref{SuppEDSymbol}.
    }
    \label{fig:vortex-ED}
\end{figure*}

In this paper, starting from the microscopic Hermitian Hamiltonian of a nonrotating quasi-2D scalar bosons in the s-wave scattering limit \cite{pethick,Castin1998}, we define the superfluid velocity via the probability current, and present the condition of the conservation of the topological charges of the vortices in general ``beyond the GPE'' case, i.e., when the mean-field limit of the Heisenberg equation of motion has a dissipative term (e.g., Refs. \cite{pitaevskii1959,Gardiner1997,Choi1998,Kasamatsu2003,Blakie2008,Damski2010,Bradley2015,Liu2020}). 
Using that condition, we show how vortices in a quasi-2D scalar BEC can be connected to the effective Maxwell's equations even beyond the PVM and without assuming a time-independent BEC number density. 
From such duality, we show that the 
damped-PVM 
\cite{Rica1990,Mehdi2023} 
can be 
alternatively induced 
and generalized beyond the PVM description and present how to calculate the temporal change of the circulation. 
Under the GPE + PVM, we show that the logarithmic vortex interaction may need correction if vortices move. We also show that one can recover previously known results when the fluctuations of the BEC number density are negligible.
For a quick reference, we present the key ideas on the duality we constructed in Fig. \ref{fig:vortex-ED}.

\section{\label{superfluidity}
Hamiltonian of quasi-2D scalar BEC and the continuity equation}

We start by showing the relation between the mean-field limit of the Heisenberg equations of motion for nonrotating quasi-2D scalar BEC on the $xy$ plane in a region $\mathcal{A}$ and vortex quantization. 
We first review the case within the GPE description and then generalize it to the case with dissipation or rotation.

Let us introduce a unit vector $\boldsymbol{e}_j$ along the $+j$ axis ($j = x, y, z$), the position vector $\boldsymbol{r} \coloneqq \sum_{j = x, y} r_j \boldsymbol{e}_j$ on the $xy$ plane, and 
$
\nabla 
\coloneqq 
\sum_{j = x, y} 
\boldsymbol{e}_j 
\partial / \partial r_j 
$. 
For later convenience, we define $\boldsymbol{e}_{\perp} \coloneqq \boldsymbol{e}_z$, 
$v \coloneqq \left\vert \boldsymbol{v} \right\vert = \sqrt{\boldsymbol{v} \cdot \boldsymbol{v}}$ for any vector $\boldsymbol{v}$, 
and 
$\left\vert \psi \right\vert \coloneqq \sqrt{\psi^{*} \psi}$ for any complex function $\psi$ with its complex conjugate being $\psi^{*}$. 
For convenience, a summary of 
the symbols we used is provided in Tables \ref{SuppSymbolDef} and \ref{SuppEDSymbol}.

\begin{table}[htbp]
\caption{
\label{SuppSymbolDef}
Definitions of symbols frequently used in this paper. 
}
\begin{ruledtabular}
\begin{tabular}{cc}
Symbol & Definition
\\
\colrule
BEC & Bose-Einstein condensates
\\
GPE & Gross-Pitaevskii equation
\\
PVM & Point-vortex model
\\
$n$D 
& 
$n$-dimensional
\\
$\mathcal{A}$ 
& 
Region where the quasi-2D BEC is
\\
$M$
& 
Mass of the boson
\\
$N$ 
& 
Number of bosons in BEC
\\
$\hat{\psi} \left( \boldsymbol{r}, t \right)$ 
& 
Bosonic field operator
\\
$\hat{n} \left( \boldsymbol{r}, t \right)$ 
& 
Number density operator [see Eqs. \eqref{SM_hatn_hatvs_def}]
\\
$\hat{\boldsymbol{J}} \left( \boldsymbol{r}, t \right)$
& 
Probability current operator [see Eqs. \eqref{SM_hatn_hatvs_def}]
\\
$\hat{A}^{\dagger}$ 
& 
Hermitian conjugate of the quantum operator $\hat{A}$
\\
$\hat{A} \pm h. c.$
& 
$\hat{A} \pm \hat{A}^{\dagger}$, where $\hat{A}$ can be any quantum operator
\\
$A$ 
& 
Quantum operator $\hat{A}$ in the mean-field limit
\\
$\psi^{*}$, $\left\vert \psi \right\vert$
& 
Complex conjugate of $\psi$, and $\left\vert \psi \right\vert \coloneqq \sqrt{\psi^{*} \psi}$
\\
$A \pm c. c.$
& 
$A \pm A^{*}$, where $A$ can be any complex function
\\
$v$ 
& 
Magnitude of the real vector $\boldsymbol{v}$ 
($
v 
\coloneqq 
\sqrt{\boldsymbol{v} \cdot \boldsymbol{v}}
$)
\\
$\Phi_K \left( \boldsymbol{r}, t \right)$ 
& 
Mean-field kinetic energy density [See Eq. \eqref{def_PhiK}]
\\
$\Phi_Q \left( \boldsymbol{r}, t \right)$ 
& 
Quantum potential [See Eq. \eqref{def_PhiQ}]
\\
$\boldsymbol{v}_s \left( \boldsymbol{r}, t \right)$ 
& 
Superfluid velocity [See Eq. \eqref{vs_def}]
\\
$\varphi \left( \boldsymbol{r}, t \right)$ 
& 
Phase of the mean-field wavefunction
\\
$\boldsymbol{f}_K \left( \boldsymbol{r}, t \right)$ 
& 
Effective kinetic force [See Eq. \eqref{def_f_K}]
\\
$U_{\rm sf} \left( \boldsymbol{r}, t \right)$ 
& 
Effective potential [See Eq. \eqref{def_Usf}]
\\
$\boldsymbol{F}_{\rm sf} \left( \boldsymbol{r}, t \right)$ 
& 
Additional force from the beyond GPE 
\\
& 
(Dissipation, rotation, etc. See Sec. \ref{BeyondGPE-superfluidity})
\\
$\boldsymbol{e}_{\perp}$ 
& 
Unit vector perpendicular to the system in $\mathcal{A}$
\\
$q_j$ 
& 
Topological charge of the vortex
\\
$\mathbb{Z}$ 
& 
Set of integers
\\
$\boldsymbol{r}_{\alpha_j}$
& 
Position of the core of the vortex [See Eq. \eqref{VortexQuantizationCond}]
\\
$N_v \left( \mathcal{M}; t \right)$ 
& 
Number of vortices in the region $\mathcal{M}$ at time $t$ 
\\
& 
[See Eq. \eqref{VortexQuantizationCond}]
\\
$\partial \mathcal{M}$ 
& 
Boundary of the region $\mathcal{M}$
\\ 
$\oint_{\partial \mathcal{M}} d \boldsymbol{l}$
& 
Closed line integral along $\partial \mathcal{M}$
\\
$\delta \left( \boldsymbol{r} \right)$
& 
Dirac delta function
\\
$\left\vert \mathcal{M} \right\vert$ 
& 
Area of the region $\mathcal{M}$
\\
$\bar{U}_{\rm sf} \left( t \right)$ 
& 
Spatial average of $U_{\rm sf} \left( \boldsymbol{r}, t \right)$ [See Eq. \eqref{def_bar_Usf}]
\\
$\theta \left( x \right)$ 
& 
Heaviside step function
\\
$\mathcal{D}^{D}_{R} \left( \boldsymbol{r} \right)$ 
& 
$D$-dimensional disk (radius $R$) centered at $\boldsymbol{r}$
\end{tabular}
\end{ruledtabular}
\end{table}

\begin{table}[htbp]
\caption{
\label{SuppEDSymbol}
Definitions of symbols for the duality to electrodynamics. 
}
\begin{ruledtabular}
\begin{tabular}{cc}
Symbol & Definition
\\
\colrule
$\rho_v \left( \boldsymbol{r}, t \right)$ 
& 
Vortex charge density [See Eq. \eqref{def_rho_v}]
\\
$\boldsymbol{E}_{\rm sf} \left( \boldsymbol{r}, t \right)$ 
& 
Effective electric field to describe the system 
\\
& 
[See Eq. \eqref{def_Esf}]
\\
$\boldsymbol{v}_{P} \left( \boldsymbol{r}, t \right)$ 
& 
Pseudo-superfluid velocity
\\
& 
[See Eqs. \eqref{def_vP} and \eqref{Poynting_S_def} for its meaning]
\\
$\epsilon_{\rm sf}$ 
& 
Effective vacuum permittivity
\\
$\boldsymbol{D}_{\rm sf} \left( \boldsymbol{r}, t \right)$ 
& 
Effective electric displacement field 
\\
& 
[See Eqs. \eqref{def_Dsf}]
\\
$\boldsymbol{P}_{\rm sf} \left( \boldsymbol{r}, t \right)$ 
& 
Effective polarization density 
[See Eqs. \eqref{def_Dsf}]
\\
$\boldsymbol{J}_{\rm sf} \left( \boldsymbol{r}, t \right)$ 
& 
Effective free electric current density 
\\
& 
[See Eq. \eqref{def_Jsf}]
\\
$\boldsymbol{H}_{\rm sf} \left( \boldsymbol{r}, t \right)$ 
& 
Effective magnetic field strength 
\\
& 
[See Eq. \eqref{def_Hsf}]
\\
$\boldsymbol{J}_{m, \rm{sf}} \left( \boldsymbol{r}, t \right)$ 
& 
Effective free magnetic current density 
\\
& 
[See Eq. \eqref{def_Jmsf}]
\\
$c_{\rm sf}$ 
& 
Effective speed of light in vacuum 
\\
& 
(Maximum speed of sound in scalar BEC)
\\
$\boldsymbol{S}_{\rm sf} \left( \boldsymbol{r}, t \right)$ 
& 
Effective Poynting vector 
[See Eq. \eqref{Poynting_S_def}]
\\
$V_{e, \rm{sf}} \left( \boldsymbol{r}, t \right)$ 
& 
Effective electric potential 
[See Eqs. \eqref{def_Ve_Asf}]
\\
$\boldsymbol{A}_{e, \rm{sf}} \left( \boldsymbol{r}, t \right)$ 
& 
Effective electric vector potential 
\\
& 
[See Eqs. \eqref{def_Ve_Asf}]
\\
$\boldsymbol{A}_{m, \rm{sf}} \left( \boldsymbol{r}, t \right)$ 
& 
Effective magnetic vector potential 
\\
& 
[See Eqs. \eqref{def_Ve_Asf}]
\\
$\boldsymbol{B}_{\rm sf} \left( \boldsymbol{r}, t \right)$ 
& 
Effective magnetic field 
\\
& 
[See Eq. \eqref{def_Bsf} and Sec. \ref{vortex_force}]
\end{tabular}
\end{ruledtabular}
\end{table}

\subsection{\label{GPE-superfluidity}
Case 1: Gross-Pitaevskii equation for nonrotating scalar BEC}

Using second quantization and the $s$-wave scattering limit, the Hamiltonian $\hat{H} \left( t \right)$ in the Heisenberg picture of a nonrotating quasi-2D scalar Bose gas on the $xy$ plane in a region $\mathcal{A}$ can be expressed as \cite{pethick,Castin1998}
\begin{eqnarray}
\hat{H} \left( t \right) 
& = & 
\int_{\mathcal{A}} d^2 r \; 
\hat{\psi}^{\dagger} \left( \boldsymbol{r}, t \right) 
\left\lbrack 
- 
\frac{\hbar^2}{2 M} \nabla^2 
+ 
V \left( \boldsymbol{r}, t \right) 
\right\rbrack 
\hat{\psi} \left( \boldsymbol{r}, t \right) 
\nonumber\\
&& 
+ 
\frac{g}{2}  
\int_{\mathcal{A}} d^2 r \; 
\hat{\psi}^{\dagger} \left( \boldsymbol{r}, t \right) 
\hat{\psi}^{\dagger} \left( \boldsymbol{r}, t \right) 
\hat{\psi} \left( \boldsymbol{r}, t \right) 
\hat{\psi} \left( \boldsymbol{r}, t \right) 
, \qquad 
\label{SM_H_given}
\end{eqnarray}
where 
$\hat{\psi} \left( \boldsymbol{r}, t \right)$ is the bosonic field operator (in Heisenberg picture) that annihilates a boson at the position $\boldsymbol{r}$ and at time $t$, 
$\hbar$ is the reduced Planck constant, 
$M$ is the mass of the boson, 
$V \left( \boldsymbol{r}, t \right)$ is a local 
external potential (with no singularity) 
satisfying 
$\left\lbrack V \left( \boldsymbol{r}, t \right), \hat{\psi} \left( \boldsymbol{r}, t \right) \right\rbrack = 0$, 
$g = 2 \sqrt{2 \pi} \hbar^2 a_B a_s / \left( M l_{\perp} \right)$ is the density-density interaction coefficient in the quasi-2D BEC \cite{pitaevskii2003bose,Bradley2015}, 
$a_B$ is the Bohr radius, $a_s$ is the $s$-wave scattering length in units of $a_B$, $l_{\perp} \coloneqq \sqrt{\hbar / \left( M \omega_{\perp} \right)}$ is the harmonic oscillator length in $z$ axis, and $\omega_{\perp}$ is the harmonic trap frequency in $z$ axis.

By defining the number density operator $\hat{n} \left( \boldsymbol{r}, t \right)$ and the probability current operator $\hat{\boldsymbol{J}} \left( \boldsymbol{r}, t \right)$ as 

\begin{eqnarray}
&& 
\hat{\boldsymbol{J}} \left( \boldsymbol{r}, t \right) 
\coloneqq 
\frac{\hbar}{2 M i} 
\left\lbrack 
\hat{\psi}^{\dagger} \left( \boldsymbol{r}, t \right) 
\nabla 
\hat{\psi} \left( \boldsymbol{r}, t \right) 
- 
h. c. 
\right\rbrack 
, 
\nonumber\\
&& 
\hat{n} \left( \boldsymbol{r}, t \right) 
\coloneqq 
\hat{\psi}^{\dagger} \left( \boldsymbol{r}, t \right) 
\hat{\psi} \left( \boldsymbol{r}, t \right) 
, 
\label{SM_hatn_hatvs_def}
\end{eqnarray}
where $\hat{A} \pm h. c. \coloneqq \hat{A} \pm \hat{A}^{\dagger}$ for any quantum operator $\hat{A}$, one can show that the integrated density operator $\int_{\mathcal{A}} d^2 r \; \hat{n} \left( \boldsymbol{r}, t \right)$ commutes with $\hat{H} \left( t \right)$ and thus 
$N \coloneqq \left\langle \int_{\mathcal{A}} d^2 r \; \hat{n} \left( \boldsymbol{r}, t \right) \right\rangle$ is constant in time $t$, with $\left\langle \hat{A} \right\rangle$ denoting the expectation value of the quantum operator $\hat{A}$. 
From the Heisenberg equations of motion, it follows that
\begin{equation}
\frac{
\partial 
\hat{n} \left( \boldsymbol{r}, t \right)}{
\partial t 
}
+ 
\nabla 
\cdot 
\hat{\boldsymbol{J}} \left( \boldsymbol{r}, t \right) 
= 
0
, 
\label{continuityEq}
\end{equation}
and

\begin{eqnarray}
\frac{
\partial 
\hat{\boldsymbol{J}} \left( \boldsymbol{r}, t \right) 
}{\partial t} 
& = & 
\frac{\hbar^2}{4 M^2} 
\left\lgroup 
\left\{ 
\nabla 
\left\lbrack 
\nabla^2  
\hat{\psi}^{\dagger} \left( \boldsymbol{r}, t \right) 
\right\rbrack 
\right\} 
\hat{\psi} \left( \boldsymbol{r}, t \right) 
+ 
h. c. 
\right\rgroup 
\nonumber\\
&& 
- 
\frac{1}{M} 
\hat{\psi}^{\dagger} \left( \boldsymbol{r}, t \right) 
\left\{ 
\nabla 
\left\lbrack 
V \left( \boldsymbol{r}, t \right) 
+ 
g 
\hat{n} \left( \boldsymbol{r}, t \right) 
\right\rbrack 
\right\} 
\hat{\psi} \left( \boldsymbol{r}, t \right) 
\nonumber\\
&& 
- 
\frac{\hbar^2}{4 M^2} 
\left\{ 
\left\lbrack 
\nabla^2  
\hat{\psi}^{\dagger} \left( \boldsymbol{r}, t \right) 
\right\rbrack 
\nabla  
\hat{\psi} \left( \boldsymbol{r}, t \right) 
+ 
h. c. 
\right\} 
. 
\label{J_op_time_deriv}
\end{eqnarray}
To simplify the problem, we focus on the mean-field limit where 
the BEC order parameter is given by 
$\psi \left( \boldsymbol{r}, t \right) 
\coloneqq 
\left\langle 
\hat{\psi} \left( \boldsymbol{r}, t \right) 
\right\rangle$, 
satisfying 
$
\psi^{\dagger} \left( \boldsymbol{r}, t \right) 
\psi \left( \boldsymbol{r}, t \right) 
= 
\psi \left( \boldsymbol{r}, t \right) 
\psi^{\dagger} \left( \boldsymbol{r}, t \right) 
$ as we consider a scalar (single-component) BEC. 
In this mean-field limit, 
one can introduce the BEC number density 
$
n \left( \boldsymbol{r}, t \right) 
\coloneqq 
\left\vert 
\psi \left( \boldsymbol{r}, t \right) 
\right\vert^2 
$ and the mean-field probability current

\begin{equation}
\boldsymbol{J} \left( \boldsymbol{r}, t \right) 
\coloneqq  
\frac{\hbar}{2 M i} 
\left\lbrack 
\psi^{*} \left( \boldsymbol{r}, t \right) 
\nabla 
\psi \left( \boldsymbol{r}, t \right) 
- 
c. c. 
\right\rbrack 
, 
\label{meanfield_J_def}
\end{equation}
where $A \pm c. c. \coloneqq A \pm A^{*}$ for any complex function $A$.

In the zero temperature limit, the integrated density is normalized as 
$\int_{\mathcal{A}} d^2 r \; 
n \left( \boldsymbol{r}, t \right) 
= 
N 
$ and the mean-field order parameter $\psi \left( \boldsymbol{r}, t \right)$ obeys the GPE \cite{Gross1961,pitaevskii1961vortex,Castin1998}

\begin{eqnarray}
i \hbar 
\frac{\partial 
\psi \left( \boldsymbol{r}, t \right) 
}{\partial t} 
= 
\left\lbrack 
- 
\frac{\hbar^2}{2 M} \nabla^2 
+ 
V \left( \boldsymbol{r}, t \right) 
+ 
g 
n \left( \boldsymbol{r}, t \right) 
\right\rbrack 
\psi \left( \boldsymbol{r}, t \right) 
, 
\nonumber\\
\label{GPE}
\end{eqnarray}
which takes the same form as the Heisenberg equation of motion for $\hat{\psi} \left( \boldsymbol{r}, t \right)$ upon the replacement $\hat{\psi} \left( \boldsymbol{r}, t \right) \rightarrow \psi \left( \boldsymbol{r}, t \right)$.
From the Hamiltonian in Eq. \eqref{SM_H_given}, we define the mean-field energy $E \left( t \right)$ of the nonrotating quasi-2D scalar BEC as

\begin{equation}
E \left( t \right) 
= 
\int_{\mathcal{A}} d^2 r \; 
\left\lbrack 
\Phi_K \left( \boldsymbol{r}, t \right) 
+ 
n \left( \boldsymbol{r}, t \right) 
V \left( \boldsymbol{r}, t \right) 
+ 
\frac{g}{2} 
n^2 \left( \boldsymbol{r}, t \right) 
\right\rbrack 
, 
\label{E_given}
\end{equation}
i.e., the energy obtained by neglecting the noncondensed bosonic particles. 
Here, 
\begin{eqnarray}
\Phi_K \left( \boldsymbol{r}, t \right) 
& \coloneqq & 
- 
\frac{\hbar^2}{4 M} 
\left\lbrack 
\psi^{*} \left( \boldsymbol{r}, t \right) 
\nabla^2 
\psi \left( \boldsymbol{r}, t \right) 
+ 
c. c. 
\right\rbrack 
\nonumber\\
& = & 
n \left( \boldsymbol{r}, t \right) 
\left\lbrack 
\Phi_Q \left( \boldsymbol{r}, t \right) 
+ 
\frac{M}{2} 
\frac{
J^2 \left( \boldsymbol{r}, t \right) 
}{
n^2 \left( \boldsymbol{r}, t \right) 
} 
\right\rbrack 
, 
\label{def_PhiK}
\end{eqnarray}
is the mean-field kinetic energy density, and

\begin{equation}
\Phi_Q \left( \boldsymbol{r}, t \right) 
\coloneqq 
- 
\frac{\hbar^2}{2 M} 
\frac{
\nabla^2 
\sqrt{ 
n \left( \boldsymbol{r}, t \right) 
}
}{
\sqrt{ 
n \left( \boldsymbol{r}, t \right) 
}
} 
, 
\label{def_PhiQ}
\end{equation}
is the quantum potential \cite{bohm1995,Wyatt05}. 
Note that $\Phi_K \left( \boldsymbol{r}, t \right)$ is well-defined even for $\psi \left( \boldsymbol{r}, t \right) = 0$, whereas $\Phi_Q \left( \boldsymbol{r}, t \right)$ is singular where $\psi \left( \boldsymbol{r}, t \right) = 0$.

From the GPE in Eq. \eqref{GPE}, it follows that

\begin{equation}
\frac{
\partial n \left( \boldsymbol{r}, t \right) 
}{\partial t} 
+ 
\nabla 
\cdot 
\boldsymbol{J} \left( \boldsymbol{r}, t \right) 
= 
0 
, 
\label{meanfield_continuity1}
\end{equation} 
and 

\begin{eqnarray}
&& 
M 
n \left( \boldsymbol{r}, t \right) 
\frac{
\partial 
\boldsymbol{J} \left( \boldsymbol{r}, t \right) 
}{\partial t} 
\nonumber\\
&& \quad 
= 
\Phi_K \left( \boldsymbol{r}, t \right) 
\nabla 
n \left( \boldsymbol{r}, t \right) 
- 
n \left( \boldsymbol{r}, t \right) 
\nabla 
\Phi_K \left( \boldsymbol{r}, t \right) 
\nonumber\\
&& \qquad 
- 
n^2 \left( \boldsymbol{r}, t \right) 
\nabla 
\left\lbrack 
V \left( \boldsymbol{r}, t \right) 
+ 
g 
n \left( \boldsymbol{r}, t \right) 
\right\rbrack 
\nonumber\\
&& \qquad 
- 
M 
\boldsymbol{J} \left( \boldsymbol{r}, t \right) 
\nabla 
\cdot 
\boldsymbol{J} \left( \boldsymbol{r}, t \right) 
. 
\label{meanfield_continuity2}
\end{eqnarray} 

Following Refs. \cite{Madelung1927,Bohm1952,Takabayasi1952}, we define the superfluid velocity $\boldsymbol{v}_s \left( \boldsymbol{r}, t \right)$ as 
\begin{equation}
n \left( \boldsymbol{r}, t \right) 
\boldsymbol{v}_s \left( \boldsymbol{r}, t \right) 
\coloneqq 
\boldsymbol{J} \left( \boldsymbol{r}, t \right) 
, 
\label{vs_def}
\end{equation}
so that Eq. \eqref{meanfield_continuity1} can be interpreted as the continuity equation for the superfluid number density $n \left( \boldsymbol{r}, t \right)$. 
However, care must be taken since $\boldsymbol{J} \left( \boldsymbol{r}, t \right)$ itself is well-defined even when  $\psi \left( \boldsymbol{r}, t \right) = 0$, whereas $\boldsymbol{v}_s \left( \boldsymbol{r}, t \right)$ is not well-defined at the location where the order parameter vanishes, $\psi \left( \boldsymbol{r}, t \right) = 0$. 
As a result of this feature, previous studies \cite{halperin1979,Ambegaokar1980,halperin1981,FISCHER1999} neglected the fluctuations on $n \left( \boldsymbol{r}, t \right)$ to avoid the singularity. However, we will show that such an approximation is not necessary.

Note that one can formally write $\boldsymbol{v}_s \left( \boldsymbol{r}, t \right) = \left( \hbar / M \right) \nabla \varphi \left( \boldsymbol{r}, t \right)$ from Eq. \eqref{vs_def} using the phase $\varphi \left( \boldsymbol{r}, t \right)$ of the mean-field wavefunction $\psi \left( \boldsymbol{r}, t \right)$ \cite{Madelung1927,Bohm1952,Takabayasi1952,pitaevskii2003bose,pethick}. However, we deliberately avoid identifying  the superfluid velocity in that way since it may mislead readers into thinking that 
$
\nabla 
\times 
\boldsymbol{v}_s \left( \boldsymbol{r}, t \right) 
= 
0
$ for any case. 
The proper treatment when one defines 
$
\boldsymbol{v}_s \left( \boldsymbol{r}, t \right) 
= 
\left( \hbar / M \right) 
\nabla 
\varphi \left( \boldsymbol{r}, t \right)
$ can be seen in, e.g., Refs. \cite{FEYNMAN1955,Anderson1966,halperin1979,Ambegaokar1980,Kleinert1995,Mazenko1997,Kleinert2008}. 
Here, we will show when $\nabla \times \boldsymbol{v}_s \left( \boldsymbol{r}, t \right)$ may not be zero. 
From Eq. \eqref{meanfield_J_def}, one can see that

\begin{equation}
\nabla 
\times 
\boldsymbol{J} \left( \boldsymbol{r}, t \right) 
= 
\frac{\hbar}{M i} 
\left\lbrack 
\nabla 
\psi^{*} \left( \boldsymbol{r}, t \right) 
\right\rbrack 
\times 
\left\lbrack 
\nabla 
\psi \left( \boldsymbol{r}, t \right) 
\right\rbrack 
, 
\label{vs_vorticity_st1}
\end{equation}
and according to Eq. \eqref{vs_def},

\begin{eqnarray}
&& 
n \left( \boldsymbol{r}, t \right) 
\nabla 
\times 
\boldsymbol{J} \left( \boldsymbol{r}, t \right) 
\nonumber\\
&& \quad 
= 
\left\lbrack 
\nabla 
n \left( \boldsymbol{r}, t \right) 
\right\rbrack 
\times 
\boldsymbol{J} \left( \boldsymbol{r}, t \right) 
+ 
n^2 \left( \boldsymbol{r}, t \right) 
\nabla 
\times 
\boldsymbol{v}_s \left( \boldsymbol{r}, t \right) 
. \qquad 
\label{vs_vorticity_st1-5}
\end{eqnarray}
Since

\begin{eqnarray}
&& 
\left\lbrack 
\nabla 
n \left( \boldsymbol{r}, t \right) 
\right\rbrack 
\times 
\boldsymbol{J} \left( \boldsymbol{r}, t \right) 
\nonumber\\
&& \quad 
= 
\frac{\hbar}{2 M i} 
\left\lbrack 
\psi \left( \boldsymbol{r}, t \right) 
\nabla 
\psi^{*} \left( \boldsymbol{r}, t \right) 
+ 
c. c.
\right\rbrack 
\nonumber\\
&& \qquad \qquad \quad 
\times 
\left\lbrack 
\psi^{*} \left( \boldsymbol{r}, t \right) 
\nabla 
\psi \left( \boldsymbol{r}, t \right) 
- 
c. c.
\right\rbrack 
\nonumber\\
&& \quad 
= 
\frac{\hbar}{M i} 
n \left( \boldsymbol{r}, t \right) 
\left\lbrack 
\nabla 
\psi^{*} \left( \boldsymbol{r}, t \right) 
\right\rbrack 
\times 
\left\lbrack 
\nabla 
\psi \left( \boldsymbol{r}, t \right) 
\right\rbrack 
,  
\qquad 
\label{vs_vorticity_st2}
\end{eqnarray}
wherever the superfluid density is finite $n \left( \boldsymbol{r}, t \right) \neq 0$, the vorticity vanishes, 
$
\nabla 
\times 
\boldsymbol{v}_s \left( \boldsymbol{r}, t \right) 
= 
0 
$. 
Conversely, the curl of $\boldsymbol{v}_s \left( \boldsymbol{r}, t \right)$ need not vanish when $n \left( \boldsymbol{r}, t \right) = 0$. 
In Sec. \ref{vortex-quantization}, we will determine the curl of $\boldsymbol{v}_s \left( \boldsymbol{r}, t \right)$ at $n \left( \boldsymbol{r}, t \right) = 0$ by using the vortex quantization.

With the definition of $\boldsymbol{v}_s \left( \boldsymbol{r}, t \right)$ in Eq. \eqref{vs_def}, Eqs. \eqref{meanfield_continuity1} and \eqref{meanfield_continuity2} can be written as 

\begin{equation}
\frac{\partial 
n \left( \boldsymbol{r}, t \right) 
}{\partial t} 
+ 
\nabla 
\cdot 
\left\lbrack 
n \left( \boldsymbol{r}, t \right) 
\boldsymbol{v}_s \left( \boldsymbol{r}, t \right) 
\right\rbrack 
= 
0 
, 
\label{meanfield_continuity_st1}
\end{equation}
and 

\begin{equation}
M 
n^2 \left( \boldsymbol{r}, t \right) 
\frac{\partial  
\boldsymbol{v}_s \left( \boldsymbol{r}, t \right) 
}{\partial t} 
= 
n^2 \left( \boldsymbol{r}, t \right) 
\left\lbrack 
\boldsymbol{f}_K \left( \boldsymbol{r}, t \right) 
- 
\nabla 
U_{\rm sf} \left( \boldsymbol{r}, t \right) 
\right\rbrack 
, 
\label{meanfield_continuity_st2}
\end{equation}
where we introduce the effective kinetic force $\boldsymbol{f}_K \left( \boldsymbol{r}, t \right)$ and the effective potential $U_{\rm sf} \left( \boldsymbol{r}, t \right)$, defined as

\begin{eqnarray}
n^2 \left( \boldsymbol{r}, t \right) 
\boldsymbol{f}_K \left( \boldsymbol{r}, t \right) 
& \coloneqq & 
\Phi_K \left( \boldsymbol{r}, t \right) 
\nabla 
n \left( \boldsymbol{r}, t \right) 
\nonumber\\
&& 
- 
n \left( \boldsymbol{r}, t \right) 
\nabla 
\Phi_K \left( \boldsymbol{r}, t \right) 
,  
\label{def_f_K}
\end{eqnarray}
and

\begin{equation}
U_{\rm sf} \left( \boldsymbol{r}, t \right) 
\coloneqq 
V \left( \boldsymbol{r}, t \right) 
+ 
g n \left( \boldsymbol{r}, t \right) 
. 
\label{def_Usf}
\end{equation}
For finite density $n \left( \boldsymbol{r}, t \right) \neq 0$, one can see that 
$
\boldsymbol{f}_K \left( \boldsymbol{r}, t \right) 
= 
- 
\nabla \left\lbrack 
\Phi_Q \left( \boldsymbol{r}, t \right) 
+ 
M v^2_s \left( \boldsymbol{r}, t \right) / 2 
\right\rbrack 
$. However, the effective kinetic force
$\boldsymbol{f}_K \left( \boldsymbol{r}, t \right)$ is singular at $n \left( \boldsymbol{r}, t \right) = 0$. 
In Sec. \ref{vortex-quantization}, we will show how to determine the curl of 
$\boldsymbol{f}_K \left( \boldsymbol{r}, t \right)$, leading to 
$
\nabla 
\times 
\boldsymbol{f}_K \left( \boldsymbol{r}, t \right) 
\neq 
0 
$ when vortices are moving.

Equation \eqref{meanfield_continuity_st1} is the continuity equation for the fluid with flow velocity $\boldsymbol{v}_s \left( \boldsymbol{r}, t \right)$. 
The absence of a drag force for $n \left( \boldsymbol{r}, t \right) \neq 0$ underlines the superfluid character, justifying to call $\boldsymbol{v}_s \left( \boldsymbol{r}, t \right)$ as the superfluid velocity. 
We kept $n^2 \left( \boldsymbol{r}, t \right)$ in Eq. \eqref{meanfield_continuity_st2} to emphasize that the superfluid velocity $\boldsymbol{v}_s \left( \boldsymbol{r}, t \right)$ and the effective kinetic force $\boldsymbol{f}_K \left( \boldsymbol{r}, t \right)$ are singular in the region where $n \left( \boldsymbol{r}, t \right) = 0$.

\subsection{\label{BeyondGPE-superfluidity}
Case 2: beyond the 
nonrotating 
Gross-Pitaevskii equation}

The Gross-Pitaevskii equation in Eq. \eqref{GPE} is obtained by neglecting the bosonic field operator of the noncondensed particles \cite{Castin1998}
 in the nonrotating case. 
However, at 
nonzero temperature, a dissipation term emerges (e.g., \cite{pitaevskii1959,Gardiner1997,Choi1998,Kasamatsu2003,Blakie2008,Damski2010,Bradley2015,Liu2020}). 
To go beyond the nonrotating dissipationless case, 
let 
us generalize Eqs. \eqref{meanfield_continuity1} and \eqref{meanfield_continuity2} as

\begin{equation}
\frac{
\partial n \left( \boldsymbol{r}, t \right) 
}{\partial t} 
+ 
\nabla 
\cdot 
\boldsymbol{J} \left( \boldsymbol{r}, t \right) 
= 
G \left( \boldsymbol{r}, t \right)  
, 
\label{meanfield_nh_continuity1}
\end{equation} 
and 

\begin{eqnarray}
&& 
M 
n \left( \boldsymbol{r}, t \right) 
\frac{
\partial 
\boldsymbol{J} \left( \boldsymbol{r}, t \right) 
}{\partial t} 
\nonumber\\
&& \quad 
= 
n^2 \left( \boldsymbol{r}, t \right) 
\left\lbrack 
\boldsymbol{f}_K \left( \boldsymbol{r}, t \right) 
- 
\nabla 
U_{\rm sf} \left( \boldsymbol{r}, t \right) 
\right\rbrack 
\nonumber\\
&& \qquad 
- 
M 
\boldsymbol{J} \left( \boldsymbol{r}, t \right) 
\nabla 
\cdot 
\boldsymbol{J} \left( \boldsymbol{r}, t \right) 
+ 
\boldsymbol{F} \left( \boldsymbol{r}, t \right) 
, \qquad 
\label{meanfield_nh_continuity2}
\end{eqnarray} 
where $G \left( \boldsymbol{r}, t \right)$ and $\boldsymbol{F} \left( \boldsymbol{r}, t \right)$ are real functions to be determined by the equation $i \hbar \partial \psi \left( \boldsymbol{r}, t \right) / \partial t$ in the model under consideration. Said differently, they are model-dependent. 
We shall focus on the specific forms of $\boldsymbol{F} \left( \boldsymbol{r} ,t \right)$ or $G \left( \boldsymbol{r}, t \right)$ for a rotating quasi-2D scalar BEC without dissipation in Sec. \ref{rotating_2D_BEC}.
Nevertheless, in Sec. \ref{VortexBEC_electrodynamics}, we will show that one can still build the effective Maxwell's equations without specifying them.

From the definition of the superfluid velocity in Eq. \eqref{vs_def}, Eq. \eqref{meanfield_nh_continuity2} can be expressed as

\begin{eqnarray}
&& 
M 
n^2 \left( \boldsymbol{r}, t \right) 
\frac{\partial 
\boldsymbol{v}_s \left( \boldsymbol{r}, t \right) 
}{\partial t} 
\nonumber\\
&& \quad 
= 
n^2 \left( \boldsymbol{r}, t \right) 
\left\lbrack 
\boldsymbol{f}_K \left( \boldsymbol{r}, t \right) 
- 
\nabla 
U_{\rm sf} \left( \boldsymbol{r}, t \right) 
\right\rbrack 
\nonumber\\
&& \qquad 
+ 
\boldsymbol{F} \left( \boldsymbol{r}, t \right) 
- 
M 
G \left( \boldsymbol{r}, t \right) 
n \left( \boldsymbol{r}, t \right) 
\boldsymbol{v}_s \left( \boldsymbol{r}, t \right) 
, \qquad 
\label{deriv_M_vs_nh_st1}
\end{eqnarray}
which implies that 
$
\boldsymbol{F} \left( \boldsymbol{r}, t \right) 
- 
M 
G \left( \boldsymbol{r}, t \right) 
n \left( \boldsymbol{r}, t \right) 
\boldsymbol{v}_s \left( \boldsymbol{r}, t \right) 
$ may be regarded as some kind of $\left( \textrm{force} \right) \times \left( \textrm{area} \right)^{-2}$ acting on the fluid at position $\boldsymbol{r}$ at time $t$.

Note that $- M G \left( \boldsymbol{r}, t \right) \boldsymbol{v}_s \left( \boldsymbol{r}, t \right) / n \left( \boldsymbol{r}, t \right)$ corresponds to the drag force, and thus, going beyond the GPE description, the system may no longer be a superfluid. However, for notational convenience, we will continue to use the word ``superfluid'', referring to $\boldsymbol{v}_s \left( \boldsymbol{r}, t \right)$ as the superfluid velocity whether the drag force exists or not. 
For notational convenience, we introduce

\begin{equation}
\boldsymbol{F}_{\rm sf} \left( \boldsymbol{r}, t \right) 
\coloneqq 
\frac{
\boldsymbol{F} \left( \boldsymbol{r}, t \right) 
- 
M 
G \left( \boldsymbol{r}, t \right) 
n \left( \boldsymbol{r}, t \right) 
\boldsymbol{v}_s \left( \boldsymbol{r}, t \right) 
}{
n^2 \left( \boldsymbol{r}, t \right) 
} 
, 
\label{def_F_sf}
\end{equation}
which might be singular in the region where $n \left( \boldsymbol{r}, t \right) = 0$. 
By using Eq. \eqref{deriv_M_vs_nh_st1}, one can determine

\begin{eqnarray}
\nabla 
\cdot 
\left\lbrack 
\boldsymbol{f}_K \left( \boldsymbol{r}, t \right) 
+ 
\boldsymbol{F}_{\rm sf} \left( \boldsymbol{r}, t \right) 
\right\rbrack 
& = & 
M 
\frac{\partial}{\partial t} 
\left\lbrack 
\nabla 
\cdot 
\boldsymbol{v}_s \left( \boldsymbol{r}, t \right) 
\right\rbrack 
\nonumber\\
&& 
+ 
\nabla^2 
U_{\rm sf} \left( \boldsymbol{r}, t \right) 
. 
\label{f_K_div}
\end{eqnarray}
It can be shown that so far, every result in this Sec. \ref{superfluidity} is also valid in 3D space. 
In the following subsection, we will consider a rotating quasi-2D scalar BEC without dissipation as an example.

\subsubsection{
\label{rotating_2D_BEC}
Example: rotating quasi-2D scalar BEC without dissipation}

Motivated by the experiment with two concentric counter-rotating superfluids \cite{Hernandez-Rajkov2024}, we will consider a rotating quasi-2D scalar BEC with angular velocity 
$
\boldsymbol{\Omega}_{\perp} \left( \boldsymbol{r}, t \right) 
= 
\Omega_{\perp} \left( \boldsymbol{r}, t \right) 
\boldsymbol{e}_{\perp}
$ that depends on position or time. 
In the rotating frame, the GPE in Eq. \eqref{GPE} becomes \cite{pitaevskii2003bose}

\begin{eqnarray}
i \hbar 
\frac{\partial 
\psi \left( \boldsymbol{r}, t \right) 
}{\partial t} 
& = & 
\left\lbrack 
- 
\frac{\hbar^2}{2 M} \nabla^2 
+ 
V \left( \boldsymbol{r}, t \right) 
+ 
g 
n \left( \boldsymbol{r}, t \right) 
\right\rbrack 
\psi \left( \boldsymbol{r}, t \right) 
\nonumber\\
&& 
- 
\boldsymbol{\Omega}_{\perp} \left( \boldsymbol{r}, t \right) 
\cdot 
\left( 
\boldsymbol{r} 
\times 
\frac{\hbar}{i} 
\nabla 
\right) 
\psi \left( \boldsymbol{r}, t \right) 
. 
\end{eqnarray}
Therefore one can see that

\begin{equation}
G \left( \boldsymbol{r}, t \right) 
= 
\left\lbrack 
\boldsymbol{\Omega}_{\perp} \left( \boldsymbol{r}, t \right) 
\times 
\boldsymbol{r} 
\right\rbrack 
\cdot 
\nabla 
n \left( \boldsymbol{r}, t \right) 
, 
\label{G_rot}
\end{equation}
and after some calculations,

\begin{eqnarray}
\boldsymbol{F} \left( \boldsymbol{r}, t \right) 
& = & 
\frac{1}{2} 
M 
G \left( \boldsymbol{r}, t \right) 
n \left( \boldsymbol{r}, t \right) 
\boldsymbol{v}_s \left( \boldsymbol{r}, t \right) 
\nonumber\\
&& 
- 
M 
n^2 \left( \boldsymbol{r}, t \right) 
\boldsymbol{\Omega}_{\perp} \left( \boldsymbol{r}, t \right) 
\times 
\boldsymbol{v}_s \left( \boldsymbol{r}, t \right) 
\nonumber\\
&& 
+ 
M 
n^2 \left( \boldsymbol{r}, t \right) 
\left\lbrack 
\nabla 
\cdot 
\boldsymbol{v}_s \left( \boldsymbol{r}, t \right) 
\right\rbrack 
\boldsymbol{\Omega}_{\perp} \left( \boldsymbol{r}, t \right) 
\times 
\boldsymbol{r} 
\nonumber\\
&& 
+ 
M 
n \left( \boldsymbol{r}, t \right) 
\left\lbrack 
\boldsymbol{v}_s \left( \boldsymbol{r}, t \right) 
\cdot 
\nabla 
n \left( \boldsymbol{r}, t \right) 
\right\rbrack 
\boldsymbol{\Omega}_{\perp} \left( \boldsymbol{r}, t \right) 
\times 
\boldsymbol{r} 
\nonumber\\
&& 
- 
M 
n^2 \left( \boldsymbol{r}, t \right) 
\boldsymbol{\Omega}_{\perp} \left( \boldsymbol{r}, t \right) 
\times 
\left\lbrack 
\left( 
\boldsymbol{r} 
\cdot 
\nabla 
\right) 
\boldsymbol{v}_s \left( \boldsymbol{r}, t \right) 
\right\rbrack 
\nonumber\\
&& 
+ 
M 
n^2 \left( \boldsymbol{r}, t \right) 
\left\lbrack 
\left( 
\boldsymbol{e}_{\perp} 
\times 
\boldsymbol{r} 
\right) 
\cdot 
\boldsymbol{v}_s \left( \boldsymbol{r}, t \right) 
\right\rbrack 
\nabla 
\Omega_{\perp} \left( \boldsymbol{r}, t \right) 
\nonumber\\
&& 
- 
\frac{1}{2} 
M 
n \left( \boldsymbol{r}, t \right) 
\left\{ 
\boldsymbol{\Omega}_{\perp} \left( \boldsymbol{r}, t \right) 
\cdot 
\left\lbrack 
\boldsymbol{r} 
\times 
\boldsymbol{v}_s \left( \boldsymbol{r}, t \right) 
\right\rbrack 
\right\} 
\nabla 
n \left( \boldsymbol{r}, t \right) 
\nonumber\\
&& 
- 
\frac{1}{2} 
M 
n \left( \boldsymbol{r}, t \right) 
\left\lbrack 
\boldsymbol{r} 
\cdot 
\boldsymbol{v}_s \left( \boldsymbol{r}, t \right) 
\right\rbrack 
\boldsymbol{\Omega}_{\perp} \left( \boldsymbol{r}, t \right) 
\times 
\nabla 
n \left( \boldsymbol{r}, t \right) 
\nonumber\\
&& 
- 
\frac{1}{2} 
M 
n \left( \boldsymbol{r}, t \right) 
\left\lbrack 
\boldsymbol{r} 
\cdot 
\nabla 
n \left( \boldsymbol{r}, t \right) 
\right\rbrack 
\boldsymbol{\Omega}_{\perp} \left( \boldsymbol{r}, t \right) 
\times 
\boldsymbol{v}_s \left( \boldsymbol{r}, t \right) 
. 
\nonumber\\
\label{F_rot}
\end{eqnarray}
Then, it can be shown that

\begin{equation}
\boldsymbol{F}_{\rm sf} \left( \boldsymbol{r}, t \right) 
= 
M 
\nabla 
\left\{ 
\boldsymbol{v}_s \left( \boldsymbol{r}, t \right) 
\cdot 
\left\lbrack 
\boldsymbol{\Omega}_{\perp} \left( \boldsymbol{r}, t \right) 
\times 
\boldsymbol{r} 
\right\rbrack 
\right\} 
, 
\label{F_G_rot}
\end{equation}
which is the generalization of Eq. (14.6) in Ref. \cite{pitaevskii2003bose} 
(see Appendix \ref{appendix_deriv_F_G_rot} for the derivation). 
When $\Omega_{\perp} \left( \boldsymbol{r}, t \right)$ is constant, Eq. \eqref{G_rot} is identical to Eq. (14.5) in Ref. \cite{pitaevskii2003bose}.

Note that 
$
\boldsymbol{F}_{\rm sf} \left( \boldsymbol{r}, t \right) 
$ is singular in the region where $n \left( \boldsymbol{r}, t \right) = 0$. 
Hence, one should be careful not to assume that 
$
\nabla 
\times 
\boldsymbol{F}_{\rm sf} \left( \boldsymbol{r}, t \right) 
= 
0 
$. 
As already advanced, we will consider that curl in the next section.

\section{\label{vortex-quantization}
Vortex quantization}

A vortex with topological charge $q_{j} \in \mathbb{Z}$ satisfies 
$\oint d \boldsymbol{l} 
\cdot 
\boldsymbol{v}_s \left( \boldsymbol{r}, t \right) 
= 
\left( 2 \pi \hbar / M \right) q_{j}$ around its core at position 
$
\boldsymbol{r} 
= 
\boldsymbol{r}_{\alpha_j}
$, where 
$\mathbb{Z}$ is the set of integers \cite{FEYNMAN1955,ginzburg1958}  (energetic and stability considerations generally restrict the values of $q_j$ to $\pm1$). 
Using Stokes' theorem, Eqs. \eqref{vs_vorticity_st1}, \eqref{vs_vorticity_st1-5}, and \eqref{vs_vorticity_st2}, 
if we assume that (within the system) the superfluid density is zero only at the core of the vortex, 
one can infer that

\begin{eqnarray}
&& 
n \left( \boldsymbol{r}_{\alpha_j} \left( t \right), t \right) 
= 
0 
. 
\nonumber\\
&& \quad 
\Rightarrow 
\nabla 
\times 
\boldsymbol{v}_s \left( \boldsymbol{r}, t \right) 
= 
\boldsymbol{e}_{\perp} 
\frac{2 \pi \hbar}{M} 
q_j 
\delta \left( \boldsymbol{r} - \boldsymbol{r}_{\alpha_j} \left( t \right) \right) 
, 
\label{vortex_constraint}
\end{eqnarray}
where 
$\delta \left( \boldsymbol{r} \right)$ is the Dirac delta function, and 
$\boldsymbol{r} = \boldsymbol{r}_{\alpha_j} \left( t \right)$ is the position of the core of the vortex with topological charge $q_j$ at time $t$. 
In BEC experiments or numerical simulations, the typical size of the 
vortex core 
is about the order of the healing length (coherence length) $\xi_h \coloneqq \hbar / \sqrt{2 M g n_m}$, where $n_m$ is the mean BEC number density \cite{ginzburg1958}.

In general, even for nonrotating BEC, the topological charge of the vortex may change over time.  
It may flip sign (e.g., from $q = 1$ to $q = -1$), turning a vortex into an antivortex, in a nonrotating quasi-2D scalar BEC under anisotropic trap potential and small $g$ 
\cite{Ripoll2001}. 
The nonconservation of the number of vortices is observed in nonrotating scalar bosons under the second-order phase transition \cite{Weiler08,Chomaz2015,Goo2021,Goo2022}. Its growth can be explained via the Kibble-Zurek mechanism \cite{delCampo2014}, while its decay can result from coarsening \cite{Biroli2010,Chesler2015}. 
Also, this change in the number of vortices is numerically shown in the stochastic GPE 
under the periodic boundary conditions 
\cite{thudiyangal2024} and in the stochastic projected GPE \cite{Liu2018}.

Considering the case of time-dependent topological charges, for multiple vortices in the nonrotating quasi-2D scalar BEC in the region $\mathcal{A}$, Eqs. \eqref{vortex_constraint} can be generalized to 

\begin{eqnarray}
&& 
\oint_{\partial \mathcal{A}} d \boldsymbol{l} 
\cdot 
\boldsymbol{v}_s \left( \boldsymbol{r}, t \right) 
= 
\frac{2 \pi \hbar}{M} 
\sum_{j = 1}^{\infty} 
q_{j} \left( t \right) 
, 
\nonumber\\
&& 
n \left( \boldsymbol{r}_{\alpha_j} \left( t \right), t \right) 
= 
0 
\quad \textrm{ for $j = 1, 2, \cdots, N_v \left( \mathcal{A}; t \right)$. }
\nonumber\\
&& \; 
\Rightarrow 
\nabla 
\times 
\boldsymbol{v}_s \left( \boldsymbol{r}, t \right) 
= 
\boldsymbol{e}_{\perp} 
\frac{2 \pi \hbar}{M} 
\sum_{j = 1}^{\infty} 
q_j \left( t \right) 
\delta \left( \boldsymbol{r} - \boldsymbol{r}_{\alpha_j} \left( t \right) \right) 
, 
\qquad 
\label{VortexQuantizationCond}
\end{eqnarray}
where 
$q_j \left( t \right) \in \mathbb{Z}$, 
$\partial \mathcal{A}$ represents the boundary of the region $\mathcal{A}$, 
$
\oint_{\partial \mathcal{A}} d \boldsymbol{l} 
\cdot 
\boldsymbol{v}_s \left( \boldsymbol{r}, t \right) 
$ represents the closed line integration of $\boldsymbol{v}_{s} \left( \boldsymbol{r}, t \right)$ along the closed curve $\partial \mathcal{A}$, 
$N_v \left( \mathcal{A}; t \right)$ is the number of vortices in the region $\mathcal{A}$ at time $t$, 
and 
$q_j \left( t \right) \neq 0$ for $j = 1, 2, \cdots, N_{v} \left( \mathcal{A}; t \right)$, whereas $q_j \left( t \right) = 0$ for $j > N_{v} \left( \mathcal{A}; t \right)$.

As we consider a quasi-2D system, 
\begin{equation}
\frac{\partial}{\partial t} 
\left\{ 
\boldsymbol{e}_{\perp} 
\cdot 
\left\lbrack 
\nabla 
\times 
\boldsymbol{v}_s \left( \boldsymbol{r}, t \right) 
\right\rbrack 
\right\} 
= 
\boldsymbol{e}_{\perp} 
\cdot 
\left\lbrack 
\nabla 
\times 
\frac{\partial 
\boldsymbol{v}_s \left( \boldsymbol{r}, t \right) 
}{\partial t} 
\right\rbrack 
, 
\label{curl_f_k_st1}
\end{equation}
and from Eq. \eqref{deriv_M_vs_nh_st1},

\begin{eqnarray}
&& 
M 
\boldsymbol{e}_{\perp} 
\cdot 
\left\lbrack 
\nabla 
\times 
\frac{\partial 
\boldsymbol{v}_s \left( \boldsymbol{r}, t \right) 
}{\partial t} 
\right\rbrack 
\nonumber\\
&& \quad 
= 
\boldsymbol{e}_{\perp} 
\cdot 
\left\{ 
\nabla 
\times 
\left\lbrack 
\boldsymbol{f}_K \left( \boldsymbol{r}, t \right) 
+ 
\boldsymbol{F}_{\rm sf} \left( \boldsymbol{r}, t \right) 
\right\rbrack 
\right\} 
. 
\label{curl_f_k_st2}
\end{eqnarray}
Thus, combining Eqs. \eqref{curl_f_k_st1} and \eqref{curl_f_k_st2} 
and using Stokes' theorem, 
it follows that

\begin{equation}
\nabla 
\times 
\left\lbrack 
\boldsymbol{f}_K \left( \boldsymbol{r}, t \right) 
+ 
\boldsymbol{F}_{\rm sf} \left( \boldsymbol{r}, t \right) 
\right\rbrack 
= 
M 
\frac{\partial}{\partial t} 
\left\lbrack 
\nabla 
\times 
\boldsymbol{v}_s \left( \boldsymbol{r}, t \right) 
\right\rbrack 
, 
\label{f_K_curl}
\end{equation}
and for special cases where 
$
n \left( \boldsymbol{r}, t \right) 
= 
0 
$ only at vortex cores,

\begin{eqnarray}
&& 
\nabla 
\times 
\left\lbrack 
\boldsymbol{f}_K \left( \boldsymbol{r}, t \right) 
+ 
\boldsymbol{F}_{\rm sf} \left( \boldsymbol{r}, t \right) 
\right\rbrack 
\nonumber\\
&& \quad 
= 
2 \pi \hbar 
\boldsymbol{e}_{\perp} 
\frac{\partial}{\partial t} 
\left\lbrack 
\sum_{j = 1}^{\infty} 
q_j \left( t \right) 
\delta \left( \boldsymbol{r} - \boldsymbol{r}_{\alpha_j} \left( t \right) \right) 
\right\rbrack 
, 
\end{eqnarray}
implying that the quantum curl dynamics can describe the system. 
Though it is not always possible to build effective Maxwell's equations in the quantum curl dynamics  \cite{Berry2023}, 
Eq. \eqref{f_K_curl} will be used in Sec. \ref{VortexBEC_electrodynamics} 
to define the effective free electric current density in the effective Maxwell's equations, and in Sec. \ref{circulation_time_change} to present how it is related to the time change of the circulation of $\boldsymbol{v}_s \left( \boldsymbol{r}, t \right)$.

\section{
\label{VortexBEC_electrodynamics}
The connection between vortices in quasi-2D BEC and electrodynamics
}

The mathematical connection between static vortices in a nonrotating quasi-2D scalar BEC and 2D electrostatics within the GPE and PVM is well established \cite{nelson2002defects}. 
In this section, we will derive the duality between vortices in a quasi-2D BEC and 2D electrodynamics in general. 
Motivated by Ref. \cite{Simula2020}, let us define the vortex charge density $\rho_v \left( \boldsymbol{r}, t \right)$ such that 

\begin{equation}
\int_{\mathcal{A}} d^2 r \; 
\rho_v \left( \boldsymbol{r}, t \right) 
\coloneqq 
\sum_{j = 1}^{N_{v} \left( \mathcal{A}; t \right)} 
q_{j} \left( t \right) 
. 
\label{def_rho_v}
\end{equation}

We define $\rho_v \left( \boldsymbol{r}, t \right)$ in this way because the curl of $\boldsymbol{v}_s \left( \boldsymbol{r}, t \right)$ need not equal the Dirac delta function if 
the superfluid density $n \left( \boldsymbol{r}, t \right)$ is zero within some finite region around the core of the vortex. 
For example, the nonlinearity of the field equation for superfluid ${}^4 \textrm{He}$ 
changes the curl of the superfluid velocity $\boldsymbol{v}_s \left( \boldsymbol{r}, t \right)$ 
to the smeared delta function (see Eq. (1.64) in Part II of Ref. \cite{Kleinert1989}). 
Note that setting 
$
\rho_v \left( \boldsymbol{r}, t \right) 
= 
\sum_{j = 1}^{\infty} 
q_j \left( t \right) 
\delta \left( 
\boldsymbol{r} 
- 
\boldsymbol{r}_{\alpha_j} \left( t \right) 
\right) 
$ 
corresponds to the PVM, so we also consider beyond the PVM.

Now, let us introduce the effective electric field $\boldsymbol{E}_{\rm sf} \left( \boldsymbol{r}, t \right)$ defined as 

\begin{equation}
\boldsymbol{E}_{\rm sf} \left( \boldsymbol{r}, t \right) 
\coloneqq 
\frac{M}{2 \pi \hbar \epsilon_{\rm sf}} 
\boldsymbol{v}_{P} \left( \boldsymbol{r}, t \right) 
\times 
\boldsymbol{e}_{\perp} 
,
\label{def_Esf}
\end{equation}
where 
$
\boldsymbol{v}_{P} \left( \boldsymbol{r}, t \right) 
$ 
satisfies 
\begin{equation}
\nabla 
\times 
\boldsymbol{v}_{P} \left( \boldsymbol{r}, t \right) 
= 
\boldsymbol{e}_{\perp} 
\frac{2 \pi \hbar}{M} 
\sum_{j = 1}^{\infty} 
q_j \left( t \right) 
\delta \left( \boldsymbol{r} - \boldsymbol{r}_{\alpha_j} \left( t \right) \right) 
, 
\label{def_vP}
\end{equation}
as if 
$
\boldsymbol{v}_{P} \left( \boldsymbol{r}, t \right) 
$ 
corresponds to the superfluid velocity in the PVM up to some irrotational vector (we will show later in Eq. \eqref{Poynting_S_def} that 
$
\boldsymbol{v}_{P} \left( \boldsymbol{r}, t \right) 
$ 
is the velocity of the vortex core), 
and the effective vacuum permittivity $\epsilon_{\rm sf}$ is some constant. Similarly, we define the effective electric displacement field $\boldsymbol{D}_{\rm sf} \left( \boldsymbol{r}, t \right)$ and the effective polarization density $\boldsymbol{P}_{\rm sf} \left( \boldsymbol{r}, t \right)$ as 
\begin{eqnarray}
\boldsymbol{D}_{\rm sf} \left( \boldsymbol{r}, t \right) 
& \coloneqq & 
\frac{M}{2 \pi \hbar} 
\boldsymbol{v}_s \left( \boldsymbol{r}, t \right) 
\times 
\boldsymbol{e}_{\perp} 
, 
\nonumber\\
\boldsymbol{P}_{\rm sf} \left( \boldsymbol{r}, t \right) 
& \coloneqq & 
\boldsymbol{D}_{\rm sf} \left( \boldsymbol{r}, t \right) 
- 
\epsilon_{\rm sf} 
\boldsymbol{E}_{\rm sf} \left( \boldsymbol{r}, t \right) 
\nonumber\\
& = & 
\frac{M}{2 \pi \hbar} 
\left\lbrack 
\boldsymbol{v}_s \left( \boldsymbol{r}, t \right) 
- 
\boldsymbol{v}_P \left( \boldsymbol{r}, t \right) 
\right\rbrack 
\times 
\boldsymbol{e}_{\perp} 
, 
\label{def_Dsf}
\end{eqnarray}
whence it follows that

\begin{equation}
\nabla 
\cdot 
\boldsymbol{D}_{\rm sf} \left( \boldsymbol{r}, t \right) 
= 
\rho_v \left( \boldsymbol{r}, t \right) 
. 
\label{sf_Maxwell_eq1}
\end{equation}
Further, note that 

\begin{equation}
\nabla \cdot \boldsymbol{P}_{\rm sf} \left( \boldsymbol{r}, t \right) 
= 
\rho_v \left( \boldsymbol{r}, t \right) 
- 
\sum_{j = 1}^{\infty} 
q_j \left( t \right) 
\delta \left( 
\boldsymbol{r} - \boldsymbol{r}_{\alpha_j} \left( t \right) 
\right) 
, 
\label{div_P_sf}
\end{equation}
represents the deviation from the PVM.

From Eq. \eqref{f_K_curl}, we introduce the effective free electric current density $
\boldsymbol{J}_{\rm sf} \left( \boldsymbol{r}, t \right) 
$ as 

\begin{equation}
\boldsymbol{J}_{\rm sf} \left( \boldsymbol{r}, t \right) 
\coloneqq 
\frac{
\boldsymbol{e}_{\perp} 
}{2 \pi \hbar} 
\times 
\left\lbrack 
\boldsymbol{f}_K \left( \boldsymbol{r}, t \right) 
+ 
\boldsymbol{F}_{\rm sf} \left( \boldsymbol{r}, t \right) 
\right\rbrack 
- 
\frac{\partial 
\boldsymbol{P}_{\rm sf} \left( \boldsymbol{r}, t \right) 
}{\partial t} 
, 
\label{def_Jsf}
\end{equation}
so that Eq. \eqref{f_K_curl} can be written as a continuity equation for the vortex charge density 

\begin{equation}
\frac{\partial 
\rho_v \left( \boldsymbol{r}, t \right) 
}{\partial t} 
+ 
\nabla 
\cdot 
\boldsymbol{J}_{\rm sf} \left( \boldsymbol{r}, t \right) 
= 
0 
.  
\label{rho_v_continuity}
\end{equation}

However,  Eq. \eqref{rho_v_continuity} does not imply vortex charge conservation; the vortex charge may not be conserved as $q \left( t \right)$ may depend on time $t$. 
This consideration, including the non-conservation of the vortex charge, is one of the main differences between our work and the previous works relying on the conservation of the vortex charge \cite{halperin1979,Mazenko1997,Ronning2023,Skogvoll2023}.
In Sec. \ref{circulation_time_change}, Eq. \eqref{rho_v_continuity} will be used to show how to calculate the change of the circulation of $\boldsymbol{v}_s \left( \boldsymbol{r}, t \right)$ in time.

From Eqs. \eqref{deriv_M_vs_nh_st1}, \eqref{def_Esf}, \eqref{def_Dsf}, and \eqref{def_Jsf}, we find

\begin{eqnarray}
\frac{\partial 
\boldsymbol{D}_{\rm sf} \left( \boldsymbol{r}, t \right) 
}{\partial t} 
& = & 
\frac{M}{2 \pi \hbar}
\frac{\partial 
\boldsymbol{v}_{P} \left( \boldsymbol{r}, t \right) 
}{\partial t} 
\times 
\boldsymbol{e}_{\perp} 
+ 
\frac{\partial 
\boldsymbol{P}_{\rm sf} \left( \boldsymbol{r}, t \right) 
}{\partial t} 
\nonumber\\
& = & 
- 
\boldsymbol{J}_{\rm sf} \left( \boldsymbol{r}, t \right) 
+ 
\nabla 
\times 
\left\lbrack 
- 
\frac{
U_{\rm sf} \left( \boldsymbol{r}, t \right) 
}{2 \pi \hbar} 
\boldsymbol{e}_{\perp} 
\right\rbrack 
. \quad 
\label{Dsf_time_deriv}
\end{eqnarray}
If $U_{\rm sf} \left( \boldsymbol{r}, t \right)$ is constant in space, Eq. \eqref{Dsf_time_deriv} cannot be distinguished from Maxwell's equations with the effective magnetic field strength $\boldsymbol{H}_{\rm sf} \left( \boldsymbol{r}, t \right) = 0$. 
Therefore let us define $\boldsymbol{H}_{\rm sf} \left( \boldsymbol{r}, t \right)$ as

\begin{equation}
\boldsymbol{H}_{\rm sf} \left( \boldsymbol{r}, t \right) 
\coloneqq 
- 
\frac{
U_{\rm sf} \left( \boldsymbol{r}, t \right) 
- 
\bar{U}_{\rm sf} \left( t \right) 
}{2 \pi \hbar} 
\boldsymbol{e}_{\perp}
, 
\label{def_Hsf}
\end{equation}
where 

\begin{equation}
\bar{U}_{\rm sf} \left( t \right) 
\coloneqq 
\frac{1}{
\left\vert \mathcal{A} \right\vert 
} 
\int_{\mathcal{A}} d^2 r \; 
U_{\rm sf} \left( \boldsymbol{r}, t \right) 
, 
\label{def_bar_Usf}
\end{equation}
is the spatial average of $U_{\rm sf} \left( \boldsymbol{r}, t \right)$ and 
$
\left\vert \mathcal{A} \right\vert 
$ 
denotes the area of the region $\mathcal{A}$. 
Then, Eq. \eqref{Dsf_time_deriv} can be written as

\begin{equation}
\nabla 
\times 
\boldsymbol{H}_{\rm sf} \left( \boldsymbol{r}, t \right) 
= 
\boldsymbol{J}_{\rm sf} \left( \boldsymbol{r}, t \right) 
+ 
\frac{\partial 
\boldsymbol{D}_{\rm sf} \left( \boldsymbol{r}, t \right) 
}{\partial t} 
. 
\label{sf_Maxwell_eq2}
\end{equation}
Additionally, as our system is in a quasi-2D,

\begin{equation}
\nabla 
\cdot 
\boldsymbol{H}_{\rm sf} \left( \boldsymbol{r}, t \right) 
= 
0 
,  
\label{sf_Maxwell_eq3}
\end{equation}
meaning that there is no effective free magnetic monopole. 

Using the definitions in Eqs. \eqref{def_Esf} and \eqref{def_Dsf}, 
\begin{equation}
\nabla 
\times 
\boldsymbol{D}_{\rm sf} \left( \boldsymbol{r}, t \right) 
= 
- 
\frac{M}{2 \pi \hbar} 
\boldsymbol{e}_{\perp} 
\nabla 
\cdot 
\boldsymbol{v}_{P} \left( \boldsymbol{r}, t \right) 
+ 
\nabla 
\times 
\boldsymbol{P}_{\rm sf} \left( \boldsymbol{r}, t \right) 
. 
\label{Dsf_curl}
\end{equation}
By defining the effective free magnetic current density $\boldsymbol{J}_{m, \rm{sf}} \left( \boldsymbol{r}, t \right)$ as

\begin{eqnarray}
\boldsymbol{J}_{m, \rm{sf}} \left( \boldsymbol{r}, t \right) 
& \coloneqq & 
c^2_{\rm sf} 
\left\lbrack 
\frac{M}{2 \pi \hbar} 
\boldsymbol{e}_{\perp} 
\nabla 
\cdot 
\boldsymbol{v}_{P} \left( \boldsymbol{r}, t \right) 
- 
\nabla 
\times 
\boldsymbol{P}_{\rm sf} \left( \boldsymbol{r}, t \right) 
\right\rbrack 
\nonumber\\
&& 
+ 
\frac{
\boldsymbol{e}_{\perp} 
}{2 \pi \hbar} 
\frac{\partial 
}{\partial t} 
\left\lbrack 
U_{\rm sf} \left( \boldsymbol{r}, t \right) 
- 
\bar{U}_{\rm sf} \left( t \right) 
\right\rbrack 
, 
\label{def_Jmsf}
\end{eqnarray}
where the effective speed of light in vacuum $c_{\rm sf}$ is some positive constant, Eq. \eqref{Dsf_curl} can be expressed as

\begin{equation}
\nabla 
\times 
\boldsymbol{D}_{\rm sf} \left( \boldsymbol{r}, t \right) 
= 
- 
\frac{1}{c^2_{\rm sf}} 
\left\lbrack 
\boldsymbol{J}_{m, \rm{sf}} \left( \boldsymbol{r}, t \right) 
+ 
\frac{\partial 
\boldsymbol{H}_{\rm sf} \left( \boldsymbol{r}, t \right) 
}{\partial t} 
\right\rbrack 
. 
\label{sf_Maxwell_eq4}
\end{equation}
From the 
Landau's criterion of superfluidity \cite{Landau1941,pitaevskii2003bose,pethick}, 
we will set $c_{\rm sf}$ to be the maximum value of the 
speed of sound 
in the scalar BEC, i.e.,  $c_s \coloneqq \sqrt{g n_{\rm max} / M}$, where $n_{\rm max}$ is the maximum value of $n \left( \boldsymbol{r}, t \right)$. 

Using Eq. \eqref{def_Jmsf}, it is straightforward to check that

\begin{equation}
\frac{\partial 
\rho_{m, \rm{sf}} \left( \boldsymbol{r}, t \right) 
}{\partial t} 
+ 
\nabla 
\cdot 
\boldsymbol{J}_{m, \rm{sf}} \left( \boldsymbol{r}, t \right) 
= 
0 
, 
\label{J_m_continuity}
\end{equation}
where the effective free magnetic charge density $\rho_{m, \rm{sf}} \left( \boldsymbol{r}, t \right)$ is zero [see Eq. \eqref{sf_Maxwell_eq3}]. 
Equation \eqref{J_m_continuity} implies the conservation of the effective free magnetic charge, so the effective free magnetic charge always remains zero.

From our effective Maxwell's equations in matter, given by Eqs. \eqref{sf_Maxwell_eq1}, \eqref{sf_Maxwell_eq2}, \eqref{sf_Maxwell_eq3}, and \eqref{sf_Maxwell_eq4}, the effective Poynting vector $\boldsymbol{S}_{\rm sf} \left( \boldsymbol{r}, t \right)$ is 

\begin{eqnarray}
\boldsymbol{S}_{\rm sf} \left( \boldsymbol{r}, t \right) 
& = & 
\boldsymbol{E}_{\rm sf} \left( \boldsymbol{r}, t \right) 
\times 
\boldsymbol{H}_{\rm sf} \left( \boldsymbol{r}, t \right) 
\nonumber\\
& = & 
\frac{
M 
\boldsymbol{v}_{P} \left( \boldsymbol{r}, t \right) 
}{\left( 2 \pi \hbar \right)^2 \epsilon_{\rm sf}} 
\left\lbrack 
U_{\rm sf} \left( \boldsymbol{r}, t \right) 
- 
\bar{U}_{\rm sf} \left( t \right) 
\right\rbrack 
, 
\label{Poynting_S_def}
\end{eqnarray}
implying that the vortex (free electric charge) moves parallel to $\boldsymbol{v}_{P} \left( \boldsymbol{r}, t \right)$, not parallel to the superfluid velocity $\boldsymbol{v}_s \left( \boldsymbol{r}, t \right)$ in general 
when the number of vortices is not conserved 
\cite{Rica1990,Mehdi2023}. 
The obtained duality is summarized in Fig. \ref{fig:vortex-ED} and 
Table \ref{table:vortex-Maxwell}.

\begin{table}[htbp]
\caption{
\label{table:vortex-Maxwell}
Effective Maxwell's equations for the vortices in the nonrotating quasi-2D scalar BEC. 
$\rho_v \left( \boldsymbol{r}, t \right)$ is the vortex charge density defined in Eq. \eqref{def_rho_v}, 
$\boldsymbol{J}_{\rm sf} \left( \boldsymbol{r}, t \right)$ is the effective free electric current density defined in Eq. \eqref{def_Jsf}, 
$\boldsymbol{J}_{m, \rm{sf}} \left( \boldsymbol{r}, t \right)$ is the effective free magnetic current density defined in Eq. \eqref{def_Jmsf}, 
and $c_{\rm sf}$ is the maximum value of the sound of speed in the scalar BEC. 
Refer to Eqs. \eqref{def_Esf}, \eqref{def_Dsf}, and \eqref{def_Hsf} for the definitions of the effective fields.  
}
\begin{ruledtabular}
\begin{tabular}{cc}
Equation & References
\\
\colrule
$
\nabla 
\cdot 
\boldsymbol{D}_{\rm sf} \left( \boldsymbol{r}, t \right) 
= 
\rho_v \left( \boldsymbol{r}, t \right) 
$
& 
Eq. \eqref{sf_Maxwell_eq1}
\\
$
\nabla 
\times 
\boldsymbol{H}_{\rm sf} \left( \boldsymbol{r}, t \right) 
= 
\boldsymbol{J}_{\rm sf} \left( \boldsymbol{r}, t \right) 
+ 
\partial 
\boldsymbol{D}_{\rm sf} \left( \boldsymbol{r}, t \right) 
/ \partial t  
$
& 
Eq. \eqref{sf_Maxwell_eq2}
\\ 
$
\nabla 
\cdot 
\boldsymbol{H}_{\rm sf} \left( \boldsymbol{r}, t \right) 
= 
0 
$
& 
Eq. \eqref{sf_Maxwell_eq3}
\\
$
c^2_{\rm sf} 
\nabla 
\times 
\boldsymbol{D}_{\rm sf} \left( \boldsymbol{r}, t \right) 
= 
- 
\boldsymbol{J}_{m, \rm{sf}} \left( \boldsymbol{r}, t \right) 
- 
\partial 
\boldsymbol{H}_{\rm sf} \left( \boldsymbol{r}, t \right) 
/ 
\partial t 
$
& 
Eq. \eqref{sf_Maxwell_eq4}
\end{tabular}
\end{ruledtabular} 
\end{table}

We emphasize that our results are of broader generality than those reported in Refs. \cite{FISCHER1999,Simula2020}, that assumed a uniform condensate density with negligible fluctuation in a nonrotating system, or Ref. \cite{johansen2024}, that assumed inhomogeneous time-independent condensate density in a nonrotating system to derive Maxwell's equations for the (2+1) dimensional superfluid universe.  
By contrast to these preceding works, the duality described in this Sec. \ref{VortexBEC_electrodynamics} can also be applied in the case of an inhomogeneous time-dependent condensate density $n \left( \boldsymbol{r}, t \right)$ in the (2+1) dimensional spacetime, even in the rotating frame [refer to Eqs. \eqref{G_rot} and \eqref{F_G_rot} for a rotating quasi-2D scalar BEC without dissipation]. 
This extension to the inhomogeneous time-dependent $n \left( \boldsymbol{r}, t \right)$ is important to study dynamics of vortices because (i) 
vortices can move, 
and (ii) $n \left( \boldsymbol{r}, t \right) = 0$ at the core of the vortex. 
Due to those properties, fluctuation of $n \left( \boldsymbol{r}, t \right)$ cannot be neglected in general.

In what follows, let us introduce the effective electric potential $V_{e, \rm{sf}} \left( \boldsymbol{r}, t \right)$, the effective electric vector potential $\boldsymbol{A}_{e, \rm{sf}} \left( \boldsymbol{r}, t \right)$, and the effective magnetic vector potential $\boldsymbol{A}_{m, \rm{sf}} \left( \boldsymbol{r}, t \right)$ such that,

\begin{eqnarray}
\boldsymbol{D}_{\rm sf} \left( \boldsymbol{r}, t \right) 
& = & 
- 
\nabla 
V_{e, \rm{sf}} \left( \boldsymbol{r}, t \right) 
- 
\frac{1}{c^2_{\rm sf}} 
\frac{\partial 
\boldsymbol{A}_{e, \rm{sf}} \left( \boldsymbol{r}, t \right) 
}{\partial t} 
\nonumber\\
&& \quad 
- 
\frac{1}{c^2_{\rm sf}} 
\nabla 
\times 
\boldsymbol{A}_{m, \rm{sf}} \left( \boldsymbol{r}, t \right) 
, 
\nonumber\\
\boldsymbol{H}_{\rm sf} \left( \boldsymbol{r}, t \right) 
& = & 
- 
\frac{1}{c^2_{\rm sf}} 
\frac{\partial 
\boldsymbol{A}_{m, \rm{sf}} \left( \boldsymbol{r}, t \right) 
}{\partial t} 
+ 
\nabla 
\times 
\boldsymbol{A}_{e, \rm{sf}} \left( \boldsymbol{r}, t \right) 
. \qquad 
\label{def_Ve_Asf}
\end{eqnarray}
One can impose the effective Lorenz gauge

\begin{equation}
\sum_{\mu = 0}^{2} 
\partial_{\mu} 
A^{\mu}_{j, \rm{sf}} \left( \boldsymbol{r}, t \right) 
= 
0 
, 
\label{Lorenz_gauge_sf}
\end{equation}
for $j = e, m$ with 
$\partial_{\mu} 
\coloneqq 
\partial / \partial x^{\mu}
$, 
$x^{\mu} 
\coloneqq 
\left( 
c_{\rm sf} t, 
\boldsymbol{r} 
\right)
$ denoting the position vector in (2+1) dimensional spacetime, 
$A^{\mu}_{e, \rm{sf}} \left( \boldsymbol{r}, t \right) 
\coloneqq 
\left( 
c_{\rm sf} 
V_{e, \rm{sf}} \left( \boldsymbol{r}, t \right), 
\boldsymbol{A}_{e, \rm{sf}} \left( \boldsymbol{r}, t \right) 
\right)
$, 
and 
$A^{\mu}_{m, \rm{sf}} \left( \boldsymbol{r}, t \right) 
\coloneqq 
\left( 
0, 
\boldsymbol{A}_{m, \rm{sf}} \left( \boldsymbol{r}, t \right) 
\right)
$.

Then, the effective Maxwell's equations in Table \ref{table:vortex-Maxwell} can be written as

\begin{equation}
\sum_{\mu = 0}^{2} 
\partial^{\mu} 
\partial_{\mu} 
A^{\nu}_{j, \rm{sf}} \left( \boldsymbol{r}, t \right) 
= 
- 
J^{\nu}_{j, \rm{sf}} \left( \boldsymbol{r}, t \right) 
, 
\label{sf_Maxwell_relativistic}
\end{equation}
where 
$
J^{\mu}_{e, \rm{sf}} \left( \boldsymbol{r}, t \right) 
\coloneqq 
\left( 
c_{\rm sf} 
\rho_v \left( \boldsymbol{r}, t \right), 
\boldsymbol{J}_{\rm sf} \left( \boldsymbol{r}, t \right) 
\right)
$, and 
$
J^{\mu}_{m, \rm{sf}} \left( \boldsymbol{r}, t \right) 
\coloneqq 
\left( 
0 
, 
\boldsymbol{J}_{m, \rm{sf}} \left( \boldsymbol{r}, t \right) 
\right)
$. 
Here, we use the Minkowski metric $\eta_{\mu \nu}$ with $\eta_{0 0} = -1$, $\eta_{j j} = 1$ for $j = 1, 2$, and $\eta_{\mu \nu} = 0$ for $\mu \neq \nu$.

Using the results in Refs. \cite{Watanabe2013,boito2020}, we find that the effective electric and magnetic vector potentials are given by  
\begin{eqnarray}
A^{\mu}_{j, \rm{sf}} \left( \boldsymbol{r}, t \right) 
& = & 
\frac{c_{\rm sf}}{2 \pi} 
\int_{\mathcal{A}} d^2 r' 
\int_{- \infty}^{\infty} d t' \; 
J^{\mu}_{j, \rm{sf}} \left( \boldsymbol{r}', t' \right) 
\nonumber\\
&& \qquad 
\times 
\frac{
\theta \left( 
c_{\rm sf} \left( t - t' \right) 
- 
\left\vert \boldsymbol{r} - \boldsymbol{r}' \right\vert 
\right) 
}{
\sqrt{
c^2_{\rm sf} \left( t - t' \right)^2 
- 
\left\vert \boldsymbol{r} - \boldsymbol{r}' \right\vert^2 
}
}
, \quad 
\label{sf_A_mu}
\end{eqnarray}
where $\theta \left( x \right)$ is the Heaviside step function with $\theta \left( x \right) = 0$ for $x < 0$ and $\theta \left( x \right) = 1$ for $x > 0$.

In principle, by solving Eq. \eqref{sf_A_mu}, one can get the effective fields and study the vortex dynamics. However, the Heaviside step function in Eq. \eqref{sf_A_mu} shows that one must take into account the history or past behavior of $J^{\mu}_{j, \rm{sf}} \left( \boldsymbol{r}, t \right)$, which is the general feature of the odd-dimensional spacetime \cite{Gurses2003} and makes it difficult to solve Eq. \eqref{sf_A_mu}.

The results in this section show that one may consider vortices in a quasi-2D scalar BEC as free electric charges in 2D matter. 
However, there is one thing missing in the duality we presented: what is the effective magnetic field $\boldsymbol{B}_{\rm sf} \left( \boldsymbol{r}, t \right)$? 
In the next section, we will define the direction of $\boldsymbol{B}_{\rm sf} \left( \boldsymbol{r}, t \right)$ from the effective Lorentz force and show that the 
damped PVM \cite{Rica1990,Mehdi2023} can be derived using that effective Lorentz force.

\section{
\label{vortex_force}
The effective force acting on vortices
}

From the duality we found above, one can see that the effective Lorentz force per unit area acting on a vortex, that is, the ``effective free electric charge,'' has two contributions
$\boldsymbol{f}_v \left( \boldsymbol{r}, t \right) 
= 
\boldsymbol{f}_1 \left( \boldsymbol{r}, t \right) 
+ 
\boldsymbol{f}_2 \left( \boldsymbol{r}, t \right) 
$, where 

\begin{eqnarray}
\boldsymbol{f}_1 \left( \boldsymbol{r}, t \right) 
& \coloneqq & 
\rho_v \left( \boldsymbol{r}, t \right) 
\boldsymbol{E}_{\rm sf} \left( \boldsymbol{r}, t \right) 
\nonumber\\
& = & 
- 
\frac{M}{2 \pi \hbar \epsilon_{\rm sf}} 
\rho_v \left( \boldsymbol{r}, t \right) 
\boldsymbol{e}_{\perp} 
\times 
\boldsymbol{v}_{P} \left( \boldsymbol{r}, t \right) 
, 
\label{Lorentz_f1_def}
\end{eqnarray}
and

\begin{eqnarray}
\boldsymbol{f}_2 \left( \boldsymbol{r}, t \right) 
& \coloneqq & 
\boldsymbol{J}_{\rm sf} \left( \boldsymbol{r}, t \right) 
\times 
\boldsymbol{B}_{\rm sf} \left( \boldsymbol{r}, t \right) 
\nonumber\\
& = & 
\left\lbrack 
\frac{
\boldsymbol{B}_{\rm sf} \left( \boldsymbol{r}, t \right) 
}{2 \pi \hbar} 
\cdot 
\boldsymbol{e}_{\perp} 
\right\rbrack 
\left\lbrack 
\boldsymbol{f}_K \left( \boldsymbol{r}, t \right) 
+ 
\boldsymbol{F}_{\rm sf} \left( \boldsymbol{r}, t \right) 
\right\rbrack 
\nonumber\\
&& 
- 
\frac{
\boldsymbol{e}_{\perp} 
}{2 \pi \hbar} 
\left\{ 
\boldsymbol{B}_{\rm sf} \left( \boldsymbol{r}, t \right) 
\cdot 
\left\lbrack 
\boldsymbol{f}_K \left( \boldsymbol{r}, t \right) 
+ 
\boldsymbol{F}_{\rm sf} \left( \boldsymbol{r}, t \right) 
\right\rbrack 
\right\} 
\nonumber\\
&& 
- 
\frac{\partial 
\boldsymbol{P}_{\rm sf} \left( \boldsymbol{r}, t \right) 
}{\partial t} 
\times 
\boldsymbol{B}_{\rm sf} \left( \boldsymbol{r}, t \right) 
. 
\label{Lorentz_f2_def}
\end{eqnarray}
As vortices are in a quasi-2D system, $\boldsymbol{B}_{\rm sf} \left( \boldsymbol{r}, t \right)$ must be parallel to $\boldsymbol{e}_{\perp}$ 
in order for  $\boldsymbol{f}_v \left( \boldsymbol{r}, t \right)$ to be in the $xy$ plane. 
However, there is no constraint on $\boldsymbol{B}_{\rm sf} \left( \boldsymbol{r}, t \right)$ to build the duality between vortices in a quasi-2D scalar BEC and electrodynamics. 
In principle, one may thus set

\begin{eqnarray}
\boldsymbol{B}_{\rm sf} \left( \boldsymbol{r}, t \right) 
& \coloneqq & 
\mu_{\rm sf} 
\left\lbrack 
\boldsymbol{H}_{\rm sf} \left( \boldsymbol{r}, t \right) 
+ 
M_{\rm sf} \left( \boldsymbol{r}, t \right) 
\boldsymbol{e}_{\perp} 
\right\rbrack 
\nonumber\\
& = & 
\mu_{\rm sf} 
\left\lbrack 
M_{\rm sf} \left( \boldsymbol{r}, t \right) 
+  
\frac{
\bar{U}_{\rm sf} \left( t \right) 
- 
U_{\rm sf} \left( \boldsymbol{r}, t \right) 
}{2 \pi \hbar} 
\right\rbrack 
\boldsymbol{e}_{\perp} 
, 
\nonumber\\
\label{def_Bsf}
\end{eqnarray}
where 
$M_{\rm sf} \left( \boldsymbol{r}, t \right) \boldsymbol{e}_{\perp}$ is the effective magnetization vector, and 
the constant $\mu_{\rm sf} \coloneqq 1 / \epsilon_{\rm sf} c^2_{\rm sf}$ is the effective vacuum permeability. 
With this choice,

\begin{widetext}
\begin{eqnarray}
\boldsymbol{f}_v \left( \boldsymbol{r}, t \right) 
& = & 
\frac{
\mu_{\rm sf} 
}{2 \pi \hbar} 
\left\lbrack 
M_{\rm sf} \left( \boldsymbol{r}, t \right) 
+ 
\frac{
\bar{U}_{\rm sf} \left( t \right) 
- 
U_{\rm sf} \left( \boldsymbol{r}, t \right) 
}{2 \pi \hbar} 
\right\rbrack 
\left\lbrack 
\boldsymbol{f}_K \left( \boldsymbol{r}, t \right) 
+ 
\boldsymbol{F}_{\rm sf} \left( \boldsymbol{r}, t \right) 
\right\rbrack 
\nonumber\\
&& 
- 
\frac{
\mu_{\rm sf} 
}{2 \pi \hbar} 
\boldsymbol{e}_{\perp} 
\times 
\left\{ 
M c^2_{\rm sf} 
\rho_v \left( \boldsymbol{r}, t \right) 
\boldsymbol{v}_{P} \left( \boldsymbol{r}, t \right) 
- 
2 \pi \hbar 
\left\lbrack 
M_{\rm sf} \left( \boldsymbol{r}, t \right) 
+ 
\frac{
\bar{U}_{\rm sf} \left( t \right) 
- 
U_{\rm sf} \left( \boldsymbol{r}, t \right) 
}{2 \pi \hbar} 
\right\rbrack 
\frac{\partial 
\boldsymbol{P}_{\rm sf} \left( t \right) 
}{\partial t} 
\right\} 
. 
\label{Lorentz_f_v}
\end{eqnarray}
\end{widetext}

Note that Eq. \eqref{Lorentz_f1_def} is consistent with the ``force per unit length on a vortex line'' in Refs. \cite{FEYNMAN1955,halperin1979}. 
Another notable thing is that Eq. \eqref{Lorentz_f_v} is similar to the 
damped-PVM \cite{Rica1990,Mehdi2023}. 
For a nonrotating quasi-2D scalar BEC confined in a boxlike trapping potential ($V \left( \boldsymbol{r}, t \right) = 0$) as considered in Ref. \cite{Mehdi2023}, 
$
M \partial \boldsymbol{v}_s \left( \boldsymbol{r}, t \right) / \partial t 
= 
\boldsymbol{f}_K \left( \boldsymbol{r}, t \right) 
+ 
\boldsymbol{F}_{\rm sf} \left( \boldsymbol{r}, t \right) 
- 
g 
\nabla 
n \left( \boldsymbol{r}, t \right) 
$, where $\boldsymbol{F}_{\rm sf} \left( \boldsymbol{r}, t \right)$ is the additional force 
arising from going beyond the nonrotating dissipationless case 
(refer to Sec. \ref{BeyondGPE-superfluidity}). 
In the GPE for a nonrotating quasi-2D BEC, minimization of the mean-field energy with respect to the phase of the mean-field $\psi \left( \boldsymbol{r}, t \right)$ gives $\nabla \cdot \boldsymbol{v}_s \left( \boldsymbol{r}, t \right) = 0$ \cite{Kleinert1989}. 
As $\nabla \cdot \boldsymbol{P}_{\rm sf} \left( \boldsymbol{r}, t \right) = 0$ in the PVM (see Eq. \eqref{div_P_sf}), for small dissipation where $\nabla \cdot \boldsymbol{v}_s \left( \boldsymbol{r}, t \right) \simeq 0$ is still a good approximation, one may set 
$
\boldsymbol{P}_{\rm sf} \left( \boldsymbol{r}, t \right) 
\simeq 
c_1 \left( t \right) 
\boldsymbol{v}_s \left( \boldsymbol{r}, t \right)
$, with $c_1 \left( t \right)$ being some function that only depends on time $t$. 
Therefore, with a suitable choice of $M_{\rm sf} \left( \boldsymbol{r}, t \right)$ and an appropriate definition of the effective vortex mass, one may derive the damped-PVM \cite{Rica1990,Mehdi2023}. 
Or one may use 
$
\boldsymbol{P}_{\rm sf} \left( \boldsymbol{r}, t \right) 
\simeq 
c_1 \left( t \right) 
\boldsymbol{v}_s \left( \boldsymbol{r}, t \right)
$ 
together with the definition of $\boldsymbol{P}_{\rm sf} \left( \boldsymbol{r}, t \right)$ in the second line in Eqs. \eqref{def_Dsf} to get 
\begin{equation}
\boldsymbol{v}_{P} \left( \boldsymbol{r}, t \right) 
\simeq 
\boldsymbol{v}_s \left( \boldsymbol{r}, t \right) 
- 
\frac{2 \pi \hbar}{M} 
c_1 \left( t \right) 
\boldsymbol{e}_{\perp} 
\times 
\boldsymbol{v}_s \left( \boldsymbol{r}, t \right) 
, 
\end{equation}
and claim from the effective Poynting vector in Eq. \eqref{Poynting_S_def} that the point vortex (free electric point charge) should move with velocity $\boldsymbol{v}_{P} \left( \boldsymbol{r}, t \right)$, which also arrives to the damped-PVM mentioned above. 
In that sense, Eq. \eqref{Lorentz_f_v} describes the generalized damped vortex model.



\section{
\label{circulation_time_change}
Temporal change of the circulation in a static area
}

Under the PVM, 
\begin{equation}
\rho_v \left( \boldsymbol{r}, t \right) 
= 
\sum_{j = 1}^{\infty} 
q_j \left( t \right) 
\delta \left( 
\boldsymbol{r} - \boldsymbol{r}_{\alpha_j} \left( t \right) 
\right) 
, 
\label{rho_v_PVM}
\end{equation}
so we may choose 
$\boldsymbol{D}_{\rm sf} \left( \boldsymbol{r}, t \right) 
= 
\epsilon_{\rm sf} 
\boldsymbol{E}_{\rm sf} \left( \boldsymbol{r}, t \right) 
$, that is, we may set $\boldsymbol{P}_{\rm sf} \left( \boldsymbol{r}, t \right) = 0$ in the PVM. 
Note that it is equivalent to set $\boldsymbol{v}_{P} \left( \boldsymbol{r}, t \right) = \boldsymbol{v}_s \left( \boldsymbol{r}, t \right)$, which is the usual choice in the PVM (e.g., \cite{Pedro2011,corso2024}). 
In fluid mechanics, for any differentiable function $Q \left( \boldsymbol{r}, t \right)$, the Reynolds transport theorem \cite{Reynolds1903,kundu2012fluid} states that 
\begin{eqnarray}
&& 
\frac{d}{d t} 
\left\lbrack 
\int_{V \left( t \right)} d^3 r \; 
Q \left( \boldsymbol{r}, t \right) 
\right\rbrack 
\nonumber\\
&& \quad 
= 
\int_{V \left( t \right)} d^3 r \; 
\left\{ 
\frac{\partial 
Q \left( \boldsymbol{r}, t \right) 
}{\partial t} 
+ 
\nabla 
\cdot 
\left\lbrack 
Q \left( \boldsymbol{r}, t \right) 
\boldsymbol{v}_{V \left( t \right)} \left( \boldsymbol{r}, t \right) 
\right\rbrack 
\right\} 
, 
\nonumber\\
\label{Reynolds_transport_theorem_3D}
\end{eqnarray}
where 
$
\boldsymbol{v}_{V \left( t \right)} \left( \boldsymbol{r}, t \right) 
$ is the velocity of the moving volume $V \left( t \right)$. 
By generalizing Eq. \eqref{Reynolds_transport_theorem_3D} to 2D systems and applying it to Eq. \eqref{rho_v_continuity}, with $\boldsymbol{P}_{\rm sf} \left( \boldsymbol{r}, t \right) = 0$ (or, equivalently, to Eq. \eqref{f_K_curl}), one finds for any static area $\mathcal{M}$ in the quasi-2D system in the region $\mathcal{A}$ that

\begin{eqnarray}
&& 
\frac{d}{d t} 
\left\lbrack 
\sum_{j = 1}^{N_v \left( \mathcal{M}; t \right)} 
q_j \left( t \right) 
\right\rbrack 
\nonumber\\
&& \quad 
= 
- 
\oint_{\partial \mathcal{M}} d l \; 
\boldsymbol{e}_n 
\cdot 
\left\lbrack 
\begin{array}{c}
\textrm{$\boldsymbol{J}_{\rm sf} \left( \boldsymbol{r}, t \right)$ with $\boldsymbol{P}_{\rm sf} \left( \boldsymbol{r}, t \right) = 0$} 
\end{array}
\right\rbrack 
\nonumber\\
&& \quad 
= 
\frac{1}{2 \pi \hbar} 
\oint_{\partial \mathcal{M}} d \boldsymbol{l} 
\cdot 
\left\lbrack 
\boldsymbol{f}_K \left( \boldsymbol{r}, t \right) 
+ 
\boldsymbol{F}_{\rm sf} \left( \boldsymbol{r}, t \right) 
\right\rbrack 
, 
\label{circulation_time_deriv}
\end{eqnarray}
where $\boldsymbol{e}_n$ is the unit outward normal vector. 
The last line in Eqs. \eqref{circulation_time_deriv} is independent of whether one uses the PVM or not, according to Eq. \eqref{f_K_curl}. 
Hence, if 
$
\boldsymbol{f}_K \left( \boldsymbol{r}, t \right) 
+ 
\boldsymbol{F}_{\rm sf} \left( \boldsymbol{r}, t \right)$ is perpendicular to $\partial \mathcal{M}$, the circulation in the region $\mathcal{M}$ does not change in time. 
Since $\boldsymbol{f}_K \left( \boldsymbol{r}, t \right)$ is related to the modulation of the BEC number density $n \left( \boldsymbol{r}, t \right)$ (see the definition in Eq. \eqref{f_K_div}), phonon emission plays a role in annihilation and creation of vortices, 
as demonstrated in Refs. \cite{Kozik2005,Kwon2021}. 
For finite density $n \left( \boldsymbol{r}, t \right) \neq 0$ in $\partial \mathcal{M}$, one can use 
$
\boldsymbol{f}_K \left( \boldsymbol{r}, t \right) 
= 
- 
\nabla \left\lbrack 
\Phi_Q \left( \boldsymbol{r}, t \right) 
+ 
M v^2_s \left( \boldsymbol{r}, t \right) / 2 
\right\rbrack 
$.

As an example, let us consider the case where only one vortex is positioned at the center of the circular quasi-2D scalar BEC with circular symmetry. We thus assume no external drag or other effect that breaks circular symmetry. 
Due to this symmetry, $\boldsymbol{f}_K \left( \boldsymbol{r}, t \right)$ is perpendicular to the boundary of any disk $\partial \mathcal{D}^2_{R} \left( 0 \right)$ with $R > 0$, where $\mathcal{D}^d_{R} \left( \boldsymbol{r}_c \right)$ denotes a $d$-dimensional disk with radius $R$ centered at $\boldsymbol{r}_c$. 
Within the GPE, 
$
\boldsymbol{F}_{\rm sf} \left( \boldsymbol{r}, t \right) 
= 
0$ for a nonrotating system, while 
$
\boldsymbol{F}_{\rm sf} \left( \boldsymbol{r}, t \right) 
= 
M 
\nabla 
\left\{ 
\boldsymbol{v}_s \left( \boldsymbol{r}, t \right) 
\cdot 
\left\lbrack 
\boldsymbol{\Omega}_{\perp} \left( \boldsymbol{r}, t \right) 
\times 
\boldsymbol{r} 
\right\rbrack 
\right\} 
$ in the case of a rotating system (see Eq. \eqref{F_G_rot}), which will also be perpendicular to $\partial \mathcal{D}^2_{R} \left( 0 \right)$ due to the symmetry. 
Therefore, the vortex will be dynamically stable within the GPE description both in the rotating and nonrotating cases. 
Conversely, for a rotating or nonrotating quasi-2D scalar circular dissipationless BEC, if the system has circular symmetry, a single vortex cannot emerge at the center if there are no vortices initially. 
This is consistent with the findings in Ref. \cite{Damski2002}, where vortex creation is described at the border of the trap.

Note that the results in this section only tells how the circulation (total topological charges of vortices) in a static area changes in time. Therefore, they do not exclude the possibility of the annihilation and creation of two vortices with opposite charges. 
In the next section, we will show that the stability of the vortex could be understood with the help of the effective Poynting vector introduced in Eq. \eqref{Poynting_S_def}.

\section{
\label{vortex_stability_Poynting}
Stability of vortices in quasi-2D scalar BEC
}
It is known that the hydrogen atom cannot be stable in classical mechanics due to the radiation. 
We will show that the vortex number conservation in the PVM for a nonrotating dissipationless quasi-2D scalar BEC in a box trap \cite{Pedro2011,Richaud2021,Richaud2023,corso2024} can be understood by using the duality we constructed in Sec. \ref{VortexBEC_electrodynamics}.

In the PVM, the vortex core size is neglected and $\boldsymbol{v}_P \left( \boldsymbol{r}, t \right) = \boldsymbol{v}_s \left( \boldsymbol{r}, t \right)$, meaning that $\boldsymbol{E}_{\rm sf} \left( \boldsymbol{r}, t \right) = \boldsymbol{D}_{\rm sf} \left( \boldsymbol{r}, t \right)$. 
Then, regardless of its charge, the vortex velocity is always perpendicular to the effective electric field $\boldsymbol{E}_{\rm sf} \left( \boldsymbol{r}, t \right)$ (see Eqs. \eqref{def_Esf}, \eqref{def_Dsf}, and \eqref{sf_Maxwell_eq1}) due to other vortices. 
Then one can conclude that the vortices cannot collide in the PVM unless one uses the damped PVM or other models that make $\boldsymbol{v}_P \left( \boldsymbol{r}, t \right) \neq \boldsymbol{v}_s \left( \boldsymbol{r}, t \right)$. 
This is consistent with the results in \cite{Pedro2011,Richaud2021,Richaud2023,corso2024}.

For a nonrotating quasi-2D scalar BEC in a box trap, $n \left( \boldsymbol{r}, t \right)$ is zero only at the boundary or at the positions of point vortices, and $n \left( \boldsymbol{r}, t \right) \simeq c$ otherwise where $c$ is some positive constant. 
Also, if there is no dissipation, $\boldsymbol{v}_s \left( \boldsymbol{r}, t \right)$ should be always parallel to the boundary [see Eq. \eqref{meanfield_continuity_st1}]. 
This makes 
$U_{\rm sf} \left( \boldsymbol{r}, t \right) \simeq \bar{U}_{\rm sf} \left( t \right)$ around any infinitesimal closed curve around the point vortex as long as no other vortices are infinitesimally close to that vortex. 
Then, in the PVM, the effective Poynting vector $\boldsymbol{S}_{\rm sf} \left( \boldsymbol{r}, t \right)$ is zero around any infinitesimal closed curve around the vortex since vortices cannot collide, meaning that there is no effective radiation and hence the vortex does not lose its energy. 
In conclusion, the PVM cannot exhibit vortex annihilation/creation in a nonrotating dissipationless quasi-2D scalar BEC in a box trap. 
Of course, this conclusion does not hold for the damped PVM or other models where $\boldsymbol{v}_P \left( \boldsymbol{r}, t \right) \neq \boldsymbol{v}_s \left( \boldsymbol{r}, t \right)$, and indeed the damped PVM can explain the vortex annihilation \cite{Mehdi2023}. 
Also, if there is dissipation, vortices may disappear at the boundary in the PVM since $\boldsymbol{v}_s \left( \boldsymbol{r}, t \right)$ is not parallel to the boundary.

From the above discussion, one can infer that the effective photons would be emitted in the annihilation of vortices. As it is known that the phonon emission plays a role in the annihilation and creation of vortices \cite{Kozik2005,Kwon2021}, the phonons in a quasi-2D scalar BEC
would behave like the effective photons.

In the next section, we discuss some implications of the duality we established using a GPE description with the PVM.

\section{
\label{Example_GPE}
Nonrotating quasi-2D scalar BEC within the Gross-Pitaevskii equation and the point-vortex model 
}

Given that $\nabla \cdot \boldsymbol{v}_s \left( \boldsymbol{r}, t \right) = 0$ in the GPE for nonrotating scalar BEC \cite{Kleinert1989}, 
we may introduce using  Eq. \eqref{f_K_div} a differentiable real function $C \left( \boldsymbol{r}, t \right)$ such that

\begin{equation}
\boldsymbol{f}_K \left( \boldsymbol{r}, t \right) 
= 
\nabla 
U_{\rm sf} \left( \boldsymbol{r}, t \right) 
+ 
\nabla 
\times 
\left\lbrack 
C \left( \boldsymbol{r}, t \right) 
\boldsymbol{e}_{\perp} 
\right\rbrack 
. 
\end{equation}
Then, using Eq. \eqref{f_K_curl}, one can see that

\begin{eqnarray}
C \left( \boldsymbol{r}, t \right) 
& = & 
- 
\hbar 
\int_{\mathcal{A}} d^2 r' \; 
\ln \left( 
\frac{
\left\vert 
\boldsymbol{r} - \boldsymbol{r}' 
\right\vert 
}{L} 
\right) 
\nonumber\\
&& \qquad 
\times 
\frac{\partial}{\partial t} 
\left\lbrack 
\sum_{j = 1}^{\infty} 
q_j \left( t \right) 
\delta \left( 
\boldsymbol{r}' 
- 
\boldsymbol{r}_{\alpha_j} \left( t \right) 
\right) 
\right\rbrack 
, \quad 
\end{eqnarray}
where $L$ is some positive constant with units of length. 
In this special case, 
by choosing $\boldsymbol{P}_{\rm sf} \left( \boldsymbol{r}, t \right) = 0$ in the PVM as we discussed below Eq. \eqref{rho_v_PVM},

\begin{eqnarray}
&& 
\boldsymbol{J}_{\rm sf} \left( \boldsymbol{r}, t \right) 
\nonumber\\
&& \quad 
= 
- 
\frac{1}{2 \pi} 
\sum_{j = 1}^{\infty} 
\frac{
\dot{q}_j \left( t \right) 
\left\lbrack 
\boldsymbol{r} 
- 
\boldsymbol{r}_{\alpha_j} \left( t \right) 
\right\rbrack 
- 
q_j \left( t \right) 
\dot{\boldsymbol{r}}_{\alpha_j} \left( t \right) 
}{
\left\vert 
\boldsymbol{r} 
- 
\boldsymbol{r}_{\alpha_j} \left( t \right) 
\right\vert^2 
} 
\nonumber\\
&& \qquad 
- 
\frac{1}{\pi} 
\sum_{j = 1}^{\infty} 
q_j \left( t \right) 
\frac{
\dot{\boldsymbol{r}}_{\alpha_j} \left( t \right) 
\cdot 
\left\lbrack 
\boldsymbol{r} 
- 
\boldsymbol{r}_{\alpha_j} \left( t \right) 
\right\rbrack 
}{
\left\vert 
\boldsymbol{r} 
- 
\boldsymbol{r}_{\alpha_j} \left( t \right) 
\right\vert^4 
} 
\left\lbrack 
\boldsymbol{r} 
- 
\boldsymbol{r}_{\alpha_j} \left( t \right) 
\right\rbrack 
\nonumber\\
&& \qquad 
+ 
\frac{1}{2 \pi \hbar} 
\boldsymbol{e}_{\perp} 
\times 
\nabla 
\left\lbrack 
V \left( \boldsymbol{r}, t \right) 
+ 
g 
n \left( \boldsymbol{r}, t \right) 
\right\rbrack 
, 
\label{Jsf_GPE_PVM}
\end{eqnarray}
where $\dot{\boldsymbol{v}} \left( t \right) \coloneqq d \boldsymbol{v} \left( t \right) / d t$ for any vector $\boldsymbol{v} \left( t \right)$. 
This explicitly shows that the vortex core movement is related to the effective free electric current density, as is expected from the duality we presented.

From the continuity equation in Eq. \eqref{meanfield_continuity_st1}, provided that $\nabla \cdot \boldsymbol{v}_s \left( \boldsymbol{r}, t \right) = 0$,

\begin{equation}
\frac{\partial 
n \left( \boldsymbol{r}, t \right) 
}{\partial t} 
= 
- 
\boldsymbol{v}_s \left( \boldsymbol{r}, t \right) 
\cdot 
\nabla 
n \left( \boldsymbol{r}, t \right) 
, 
\label{meanfield_continuity_material}
\end{equation}
and 
since we set $\boldsymbol{P}_{\rm sf} \left( \boldsymbol{r}, t \right) = 0$, 
\begin{eqnarray}
&& 
\boldsymbol{J}_{m, \rm{sf}} \left( \boldsymbol{r}, t \right) 
\nonumber\\
&& \quad 
= 
\frac{
\boldsymbol{e}_{\perp} 
}{2 \pi \hbar} 
\frac{\partial}{\partial t} 
\left\lbrack 
V \left( \boldsymbol{r}, t \right) 
- 
\frac{1}{
\left\vert \mathcal{A} \right\vert 
} 
\int_{\mathcal{A}} d^2 r_1 \; 
V \left( \boldsymbol{r}_1, t \right) 
\right\rbrack 
\nonumber\\
&& \qquad 
+ 
g 
\frac{
\boldsymbol{e}_{\perp} 
}{2 \pi \hbar} 
\frac{\partial}{\partial t} 
\left\lbrack 
n \left( \boldsymbol{r}, t \right) 
- 
\frac{1}{
\left\vert \mathcal{A} \right\vert 
} 
\int_{\mathcal{A}} d^2 r_1 \; 
n \left( \boldsymbol{r}_1, t \right) 
\right\rbrack 
. \qquad 
\label{Jmsf_GPE_PVM}
\end{eqnarray}
This implies that the description can be simplified if the external potential $V \left( \boldsymbol{r}, t \right)$ and the BEC number density $n \left( \boldsymbol{r}, t \right)$ are constant in time, or their variation can be neglected. This motivates the following example.

\subsection{
Homogeneous BEC number density limit
}

Let us ignore the spatial fluctuation of $n \left( \boldsymbol{r}, t \right)$ for simplicity. 
Of course, this limit is valid only for vanishing healing length $\xi_h \rightarrow 0$; else, the spatial fluctuation of $n \left( \boldsymbol{r}, t \right)$ cannot be neglected since $n \left( \boldsymbol{r}, t \right) = 0$ at cores of vortices. 
This neglect of the size of the vortex is another assumption in the PVM. 
As we already discussed in Sec. \ref{vortex_stability_Poynting}, the effective Poynting vector is approximately zero in this limit, reassuring the known results that 
the vortices in nonrotating quasi-2D scalar BEC cannot collide in the PVM, thus leading to the vortex number conservation in the GPE + PVM description \cite{Richaud2021,Richaud2023}.

If vortices do not move, 
$
\partial 
\rho_v \left( \boldsymbol{r}, t \right) 
/ 
\partial t = 0 
$ and 
$\boldsymbol{D}_{\rm sf} \left( \boldsymbol{r}, t \right) = - \nabla V_{e, \rm{sf}} \left( \boldsymbol{r}, t \right)$, 
since $\boldsymbol{J}_{\rm sf} \left( \boldsymbol{r}, t \right) \simeq 0$ from Eq. \eqref{Jsf_GPE_PVM}. 
By using the 2D Green's function with a static source \cite{Watanabe2013},

\begin{eqnarray}
V_{e, \rm{sf}} \left( \boldsymbol{r}, t \right) 
& = & 
\int_{\mathcal{A}} d^2 r' \; 
\frac{1}{2 \pi} 
\rho_v \left( \boldsymbol{r}' \right) 
\ln \left( 
\frac{L}{
\left\vert 
\boldsymbol{r} - \boldsymbol{r}' 
\right\vert 
}
\right) 
\nonumber\\
& = & 
\sum_{j = 1}^{N_v \left( \mathcal{A} \right)}  
\frac{q_j}{2 \pi} 
\ln \left( 
\frac{L}{
\left\vert 
\boldsymbol{r} - \boldsymbol{r}_{\alpha_j} 
\right\vert 
}
\right) 
. 
\label{V_e_sf_calc_st1}
\end{eqnarray}
As an upshot, one can recover the well-known logarithmic vortex interaction energy \cite{FEYNMAN1955,ginzburg1958,pethick}. 

Now, let us consider a more general case in which vortices are created at $t = 0$ and vortices move after creation, i.e., 
$\rho_v \left( \boldsymbol{r}, t \right) = \sum_{j = 1}^{N_v \left( \mathcal{A} \right)} q_j \delta \left( \boldsymbol{r} - \boldsymbol{r}_{\alpha_j} \left( t \right) \right)$ for $t \ge 0$ and zero otherwise. 
In the near-field approximation \cite{Lazar2013,Tsaloukidis2024} where the retardation is negligible,

\begin{eqnarray}
&& 
\int_{0}^{\infty} d t' \; 
\frac{
\theta \left( 
c_{\rm sf} \left( t - t' \right) 
- 
\left\vert \boldsymbol{r} - \boldsymbol{r}_{\alpha_j} \left( t' \right) \right\vert 
\right) 
}{
\sqrt{
c^2_{\rm sf} \left( t - t' \right)^2 
- 
\left\vert \boldsymbol{r} - \boldsymbol{r}_{\alpha_j} \left( t' \right) \right\vert^2 
}
}
\nonumber\\
&& \quad 
\simeq 
\int_{0}^{
t 
} d t' \; 
\frac{
\theta \left( 
c_{\rm sf} \left( t - t' \right) 
- 
d_j \left( t \right) 
\right) 
}{
\sqrt{
c^2_{\rm sf} \left( t - t' \right)^2 
- 
d^2_j \left( t \right) 
}
} 
\nonumber\\
&& \quad 
= 
\theta \left( 
c_{\rm sf} t - d_j \left( t \right) 
\right) 
\int_{d_j \left( t \right) / c_{\rm sf}}^{t} 
\frac{
d \tau 
}{
\sqrt{
\left( c_{\rm sf} \tau \right)^2 
- 
d^2_j \left( t \right) 
}
}
, \qquad 
\end{eqnarray}
where $
d_j \left( t \right) 
\coloneqq 
\left\vert 
\boldsymbol{r} - \boldsymbol{r}_{\alpha_j} \left( t \right) 
\right\vert 
$ and $\tau \coloneqq t - t'$. 
From this result,

\begin{eqnarray}
&& 
V_{e; j} \left( \boldsymbol{r}, t \right) 
\nonumber\\
&& \quad 
\coloneqq 
\frac{c_{\rm sf}}{2 \pi} 
q_j 
\int_{0}^{\infty} d t' \; 
\frac{
\theta \left( 
c_{\rm sf} \left( t - t' \right) 
- 
\left\vert \boldsymbol{r} - \boldsymbol{r}_{\alpha_j} \left( t' \right) \right\vert 
\right) 
}{
\sqrt{
c^2_{\rm sf} \left( t - t' \right)^2 
- 
\left\vert \boldsymbol{r} - \boldsymbol{r}_{\alpha_j} \left( t' \right) \right\vert^2 
}
}
\nonumber\\
&& \quad 
\simeq  
\frac{q_j}{2 \pi} 
\ln \left\lbrack 
\frac{
c_{\rm sf} t 
+ 
\sqrt{
\left( c_{\rm sf} t \right)^2 
- 
\left\vert 
\boldsymbol{r} - \boldsymbol{r}_{\alpha_j} \left( t \right) 
\right\vert^2 
}
}{
\left\vert 
\boldsymbol{r} - \boldsymbol{r}_{\alpha_j} \left( t \right) 
\right\vert 
}
\right\rbrack 
\nonumber\\
&& \qquad \quad 
\times 
\theta \left( 
c_{\rm sf} t 
- 
\left\vert 
\boldsymbol{r} - \boldsymbol{r}_{\alpha_j} \left( t \right) 
\right\vert 
\right) 
, 
\label{Example_V_e_j_calc}
\end{eqnarray}
in the near-field approximation. 
Note that $V_{e, \rm{sf}} \left( \boldsymbol{r}, t \right) = \sum_{j = 1}^{\infty} V_{e; j} \left( \boldsymbol{r}, t \right)$, so $V_{e; j} \left( \boldsymbol{r}, t \right)$ is the effective electric potential due to the vortex with topological charge $q_j$ whose core is at $\boldsymbol{r}_{\alpha_j} \left( t \right)$ at time $t \ge 0$. 
Since $V_{e; j} \left( \boldsymbol{r}, t \right) = 0$ for $c_{\rm sf} t < \left\vert \boldsymbol{r} - \boldsymbol{r}_{\alpha_j} \left( t \right) \right\vert$, we will focus on the case where $c_{\rm sf} t > \left\vert \boldsymbol{r} - \boldsymbol{r}_{\alpha_j} \left( t \right) \right\vert$. 
Note that

\begin{equation}
V_{e; j} \left( \boldsymbol{r}, t \right) 
\simeq 
\frac{q_j}{2 \pi} 
\ln \left( 
\frac{
2 c_{\rm sf} t 
}{
\left\vert 
\boldsymbol{r} - \boldsymbol{r}_{\alpha_j} \left( t \right) 
\right\vert 
}
\right) 
+ 
\ord \left( 
\frac{
\left\vert 
\boldsymbol{r} - \boldsymbol{r}_{\alpha_j} \left( t \right) 
\right\vert^2 
}{\left( c_{\rm sf} t \right)^2} 
\right) 
. 
\label{V_e_approx}
\end{equation}

Let us assume there is a single moving vortex with charge $q_1$ at $\boldsymbol{r}_{\alpha_1} \left( t \right)$ for $t \ge 0$. 
As we obtain $V_{e, \rm{sf}} \left( \boldsymbol{r}, t \right)$ in the near-field approximation, 
for $c_{\rm sf} t > \left\vert \boldsymbol{r} - \boldsymbol{r}_{\alpha_1} \left( t \right) \right\vert$, 
the approximate superfluid velocity $\boldsymbol{v}^{\left( 0 \right)}_s \left( \boldsymbol{r}, t \right)$ when neglecting $\boldsymbol{J}_{\rm sf} \left( \boldsymbol{r}, t \right)$ is

\begin{equation}
\boldsymbol{v}^{\left( 0 \right)}_s \left( \boldsymbol{r}, t \right) 
\coloneqq 
- 
\frac{2 \pi \hbar}{M} 
\boldsymbol{e}_{\perp} 
\times 
\nabla 
V_{e, \rm{sf}} \left( \boldsymbol{r}, t \right) 
= 
\sum_{j = 1}^{2} 
\boldsymbol{v}^{\left( 0 \right)}_{s; j} \left( \boldsymbol{r}, t \right) 
, 
\end{equation}
where

\begin{equation}
\boldsymbol{v}^{\left( 0 \right)}_{s; 1} \left( \boldsymbol{r}, t \right) 
\simeq 
q_1 
\frac{\hbar}{M} 
\boldsymbol{e}_{\perp} 
\times 
\frac{
\boldsymbol{r} - \boldsymbol{r}_{\alpha_1} \left( t \right) 
}{
\left\vert 
\boldsymbol{r} - \boldsymbol{r}_{\alpha_1} \left( t \right) 
\right\vert^2 
} 
, 
\label{v_sf_GPE_PVM_1}
\end{equation}
and

\begin{eqnarray}
\boldsymbol{v}^{\left( 0 \right)}_{s; 2} \left( \boldsymbol{r}, t \right) 
& \simeq & 
q_1 
\frac{\hbar}{M} 
\boldsymbol{e}_{\perp} 
\times 
\frac{
\boldsymbol{r} - \boldsymbol{r}_{\alpha_1} \left( t \right) 
}{
c_{\rm sf} t 
+ 
\sqrt{
\left( c_{\rm sf} t \right)^2 
- 
\left\vert 
\boldsymbol{r} - \boldsymbol{r}_{\alpha_1} \left( t \right) 
\right\vert^2 
}
} 
\nonumber\\
&& 
\times 
\frac{1}{
\sqrt{
\left( c_{\rm sf} t \right)^2 
- 
\left\vert 
\boldsymbol{r} - \boldsymbol{r}_{\alpha_1} \left( t \right) 
\right\vert^2 
}
} 
,  
\label{v_sf_GPE_PVM_2}
\end{eqnarray}
in the near-field approximation.

The first component, $\boldsymbol{v}^{\left( 0 \right)}_{s; 1} \left( \boldsymbol{r}, t \right)$ in Eq. \eqref{v_sf_GPE_PVM_1}, is the simplest extension of the superfluid velocity due to the single static vortex whose core is at $\boldsymbol{r}_{\alpha_1}$. 
However, that is not enough for a moving vortex, as one may infer from the electric field of the moving particle (Jefimenko's equations \cite{jefimenko1966,Griffiths_2017}). 
Note that

\begin{equation}
\boldsymbol{v}^{\left( 0 \right)}_{s; 2} \left( \boldsymbol{r}, t \right) 
\simeq 
q_1 
\frac{\hbar}{M} 
\boldsymbol{e}_{\perp} 
\times 
\frac{
\boldsymbol{r} - \boldsymbol{r}_{\alpha_1} \left( t \right) 
}{
2 \left( c_{\rm sf} t \right)^2 
} 
+ 
\ord \left( 
\frac{
\left\vert 
\boldsymbol{r} - \boldsymbol{r}_{\alpha_1} \left( t \right) 
\right\vert^2 
}{\left( c_{\rm sf} t \right)^2} 
\right) 
. 
\label{v_sf_GPE_PVM_2_approx}
\end{equation}
One can check that $
\nabla 
\times 
\left\{ 
\boldsymbol{e}_{\perp} 
\times 
\left\lbrack 
\boldsymbol{r} - \boldsymbol{r}_{\alpha_1} \left( t \right) 
\right\rbrack 
\right\} 
= 
2 
\boldsymbol{e}_{\perp}
$ as we consider a quasi-2D system. However, Eqs. \eqref{v_sf_GPE_PVM_2} and \eqref{v_sf_GPE_PVM_2_approx} are valid only in the near-field approximation and thus the closed line integral of Eq. \eqref{v_sf_GPE_PVM_2_approx} around the core of the vortex at $\boldsymbol{r}_{\alpha_1} \left( t \right)$ is of order
$
\ord \left( 
\left\vert 
\boldsymbol{r} - \boldsymbol{r}_{\alpha_1} \left( t \right) 
\right\vert^2 
/ 
\left( c_{\rm sf} t \right)^2 
\right) 
$
, which is small since $c_s t > \left\vert \boldsymbol{r} - \boldsymbol{r}_{\alpha_1} \left( t \right) \right\vert$.

With the above results, 

\begin{eqnarray}
&& 
\nabla 
\left\lbrack 
\boldsymbol{v}^{\left( 0 \right)}_s \left( \boldsymbol{r}, t \right) 
\cdot 
\boldsymbol{v}^{\left( 0 \right)}_s \left( \boldsymbol{r}, t \right) 
\right\rbrack 
\nonumber\\
&& \quad 
\simeq 
- 
2 
\left( 
q_1 
\frac{\hbar}{M} 
\right)^2 
\frac{
\boldsymbol{r} - \boldsymbol{r}_{\alpha_1} \left( t \right) 
}{
\left\vert 
\boldsymbol{r} - \boldsymbol{r}_{\alpha_1} \left( t \right) 
\right\vert^4 
} 
\theta \left( 
c_{\rm sf} t 
- 
\left\vert 
\boldsymbol{r} - \boldsymbol{r}_{\alpha_1} \left( t \right) 
\right\vert 
\right) 
\nonumber\\
&& \qquad 
+ 
\ord \left( 
\frac{
\left\vert 
\boldsymbol{r} - \boldsymbol{r}_{\alpha_1} \left( t \right) 
\right\vert^2 
}{\left( c_{\rm sf} t \right)^2} 
\right) 
, \qquad 
\end{eqnarray}
in the near-field approximation, showing that $\boldsymbol{J}_{\rm sf} \left( \boldsymbol{r}, t \right) \neq 0$. 
The correction due to the nonzero $\boldsymbol{J}_{\rm sf} \left( \boldsymbol{r}, t \right)$ is of order $
\ord \left( 
\left\vert 
\boldsymbol{r} - \boldsymbol{r}_{\alpha_1} \left( t \right) 
\right\vert^2 
/ 
\left( c_{\rm sf} t \right)^2 
\right) 
$ relative to $\boldsymbol{v}^{\left( 0 \right)}_{s; 1} \left( \boldsymbol{r}, t \right)$, but further evaluation needs numerical calculations to solve 
Eq. \eqref{sf_A_mu} under the specific system in the region $\mathcal{A}$ one studies. 
One may consider it as solving the effective Li{\'e}nard–Wiechert potentials in (2+1) dimensional spacetime since vortices in the PVM can be regarded as free electric point charges, according to the duality we described. 
Given that the main subject of this paper is to present the duality between vortices in a quasi-2D scalar BEC and electrodynamics, we do not discuss this further.

\subsection{
\label{Example_GPE_static}
Relation to 2D Coulomb gas and vortex spacing distribution
}

As we showed in Eq. \eqref{V_e_sf_calc_st1}, the static vortices in the nonrotating quasi-2D BEC in the GPE + PVM description can be mapped to the 2D Coulomb gas. 
Using the duality between the 2D Coulomb gas and the (1+1) dimensional sine-Gordon model \cite{FrohlichPRL1975,Frohlich1976}, the equivalence between the sine-Gordon model and the massive Thirring model \cite{Coleman1975}, and the work in Ref. \cite{Samuel1978} that connected between the Berezinskii-Kosterlitz-Thouless (BKT) transition \cite{berezinskii1972,Kosterlitz_1973} and the 2D Coulomb gas, one can infer that the BKT transition may happen in the nonrotating quasi-2D BEC with vortices. 
Using the results in Ref. \cite{Samaj2DCoulomb}, if the topological charge of the vortex is $\pm Q$ and the total charge is zero, it can be shown that the BKT transition occurs when the effective temperature $T_{\rm eff}$ of the system is at $T_c$, which takes the value

\begin{equation}
T_c 
= 
\frac{
n \pi \hbar^2 Q^2
}{
2 M k_B 
}
, 
\end{equation}
in the GPE + PVM description.

Note that this effective temperature $T_{\rm eff}$ may not be directly related to the temperature of the BEC system. 
The GPE is the zero temperature limit of the Heisenberg equation of motion for the bosonic field operator $\hat{\psi} \left( \boldsymbol{r}, t \right)$. 
Nevertheless, if there is no vortex dipole in the simulation using the GPE, $T_{\rm eff} > T_c$. Otherwise, $T_{\rm eff} < T_c$.

Another notable connection is that, in a newborn superfluid, the early vortex spacing distribution closely follows the Poisson point process (PPP) in the PVM with a density predicted by the Kibble-Zurek mechanism (KZM) \cite{delcampo22,thudiyangal2024}. Further, the spacing distribution of the 2D Coulomb gas also follows a PPP when the effective temperature is infinite \cite{Akemann2022,Massaro2024}. 
Assuming that the initial movement of the vortices can be neglected, the topological charge of each vortex is conserved initially, and the deviation from the GPE is negligible, vortices can be mapped to the 2D Coulomb gas when the measurement is done soon after the vortex creation. 
Then, we may understand the similarity between Refs. \cite{thudiyangal2024} and \cite{Akemann2022,Massaro2024} with the help of the duality discussed in this paper: the initial effective temperature of the vortices is very high when the vortices are created via KZM (no vortex dipole exists initially), and they cool down as time goes on until the duality between vortices and 2D Coulomb gas is broken, given that the motion of vortices cannot be neglected as time goes on. 
After that, vortex-antivortex annihilation occurs instead of forming vortex dipoles since vortices can no longer be mapped to the 2D Coulomb gas. One should solve Eq. \eqref{sf_A_mu} to find corrections beyond the 2D Coulomb gas.  

\section{Conclusion}

The description of vortices in two-dimensional systems as a Coulomb gas has a fruitful history.
In this work, we have provided a description of vortices in a quasi-2D scalar BEC in terms of 2D electrodynamics. Such duality goes beyond the previous findings not only by deriving the analog of Maxwell's equations that account for inhomogeneous time-dependent BEC with and without dissipation but also by considering the superfluid rotation.

We have elucidated how to map the vortices in a quasi-2D scalar BEC to Maxwell's equations in (2+1) dimensional spacetime. 
Such formulation may find applications in the study of nonequilibrium BEC proliferated by vortices, with applications to the study of vortex patterns, including clustering and melting \cite{Gauthier2019,Neely2024}, quantum turbulence \cite{Tsatsos2016,Reeves2022}, stochastic geometry of quantum matter \cite{delcampo22,thudiyangal2024,Massaro2024}, and vortex pattern detection via a quantum dynamical microscope, that relies on controlled expansions realizing a shortcut to adiabaticity to scale up the superfluid cloud \cite{delcampo11,delcampo13,Papoular15}. Likewise, our findings can be applied to the dynamics of phase transitions \cite{Weiler08,Chomaz2015,Navon2015,Goo2021,Goo2022, Kim2023}, including the Kibble-Zurek mechanism \cite{DRP11,delCampo2014}, the generalizations to account for the universal statistics of defects \cite{GomezRuiz20,delCampo2021}, and fast-quench universality \cite{Zeng2023,Xia2024}. 

An interesting prospect is the generalization of our results to BEC characterized by higher-order nonlinearities (e.g., due to losses and confinement), dipolar interactions \cite{Lahaye2009}, and spinor degrees of freedom \cite{Kawaguchi2012}, 
where the mean-field wavefunction 
$
\psi \left( \boldsymbol{r}, t \right) 
= 
\left\lbrack 
\begin{array}{ccc} 
\psi_{f} \left( \boldsymbol{r}, t \right) 
& 
\cdots 
& 
\psi_{- f} \left( \boldsymbol{r}, t \right) 
\end{array} 
\right\rbrack^T  
$ 
for a spin-$f$ system does not commute with its Hermitian conjugate $\psi^{\dagger} \left( \boldsymbol{r}, t \right)$ in general. 

Beyond the realm of ultracold gases, our findings can be applied and generalized for polaritonic BEC and quantum fluids of light \cite{Carusotto13}.  
The extending of our results beyond the quasi-2D case is also an open problem. For example, in 3D, one can infer from Eqs. \eqref{sf_Maxwell_eq2} and \eqref{sf_Maxwell_eq3} that there would be effective free magnetic charge since $\boldsymbol{e}_{\perp}$ would correspond to the unit vector along the vortex line, which is both space- and time-dependent as the vortex line can be open or closed \cite{Kleinert1995,Weiler08,Fujiyama2009,Adachi2010,Serafini2015}. 
This effective free magnetic charge would affect the effective magnetic field strength $\boldsymbol{H}_{\rm sf} \left( \boldsymbol{r}, t \right)$, introducing additional terms to the vortex line interaction in Refs. \cite{Kleinert1989,Kleinert2008} that neglected vortex core regions.

One may also wonder whether other electromagnetic dualities exist to describe the different types of topological defects that are classified by homotopy theory \cite{Kibble1976,Mermin1979,Manton2004}. Beyond vortices, are there electromagnetic dualities valid for domain walls, monopoles, textures, and skyrmions? Can a unified duality valid for any type of topological defect be conceived?
Recently, a similar duality to the one we have reported has been introduced for defects in crystalline solids in the Hermitian case of elastic media \cite{Tsaloukidis2024}. This suggests that it might be possible to build a duality between topological defects and Maxwell's equations in other systems.

\begin{acknowledgments}
It is a pleasure to thank 
Michael V. Berry, 
Ashton Bradley, 
Bogdan Damski, \'Iñigo L. Egusquiza, 
Uwe R. Fischer, 
{\'E}tienne Fodor, Andr\'as Grabarits, 
Woo Jin Kwon, 
Matteo Massaro, Kazutaka Takahashi, 
Mithun Thudiyangal, 
and Masahito Ueda for insightful comments and discussions.
The authors acknowledge financial support from the National Research Fund of Luxembourg under Grant No. C22/MS/17132060/BeyondKZM. 
For the purpose of open access, the author has applied a Creative Commons Attribution 4.0 International (CC BY 4.0) license to any Author Accepted Manuscript version arising from this submission.
\end{acknowledgments}

\appendix

\section{
\label{appendix_deriv_F_G_rot}
Derivation of Eq. \eqref{F_G_rot}}

In a quasi-2D system with 
$
\boldsymbol{\Omega}_{\perp} \left( \boldsymbol{r}, t \right) 
= 
\Omega_{\perp} \left( \boldsymbol{r}, t \right) 
\boldsymbol{e}_{\perp} 
$,

\begin{eqnarray}
&& 
\nabla 
\left\{ 
\boldsymbol{v}_s \left( \boldsymbol{r}, t \right) 
\cdot 
\left\lbrack 
\boldsymbol{\Omega}_{\perp} \left( \boldsymbol{r}, t \right) 
\times 
\boldsymbol{r} 
\right\rbrack 
\right\} 
\nonumber\\
&& \quad 
= 
- 
\boldsymbol{r} 
\times 
\boldsymbol{e}_{\perp} 
\left\lbrack 
\boldsymbol{v}_s \left( \boldsymbol{r}, t \right) 
\cdot 
\nabla 
\Omega_{\perp} \left( \boldsymbol{r}, t \right) 
\right\rbrack 
\nonumber\\
&& \qquad 
- 
\boldsymbol{r} 
\times 
\boldsymbol{\Omega}_{\perp} \left( \boldsymbol{r}, t \right) 
\left\lbrack 
\nabla 
\cdot 
\boldsymbol{v}_s \left( \boldsymbol{r}, t \right) 
\right\rbrack 
+ 
\boldsymbol{v}_s \left( \boldsymbol{r}, t \right) 
\times 
\boldsymbol{\Omega}_{\perp} \left( \boldsymbol{r}, t \right) 
\nonumber\\
&& \qquad 
+ 
\left\lbrack 
\left( 
\boldsymbol{r} 
\cdot 
\nabla 
\right) 
\boldsymbol{v}_s \left( \boldsymbol{r}, t \right) 
\right\rbrack 
\times 
\boldsymbol{\Omega}_{\perp} \left( \boldsymbol{r}, t \right) 
\nonumber\\
&& \qquad 
+ 
\boldsymbol{v}_s \left( \boldsymbol{r}, t \right) 
\times 
\left\lbrack 
\left( 
\boldsymbol{r} 
\cdot 
\nabla 
\right) 
\boldsymbol{\Omega}_{\perp} \left( \boldsymbol{r}, t \right) 
\right\rbrack 
. 
\end{eqnarray}
From Eqs. \eqref{G_rot} and \eqref{F_rot},

\begin{eqnarray}
&& 
\boldsymbol{F} \left( \boldsymbol{r}, t \right) 
- 
M 
G \left( \boldsymbol{r}, t \right) 
n \left( \boldsymbol{r}, t \right) 
\boldsymbol{v}_s \left( \boldsymbol{r}, t \right) 
\nonumber\\
&& \quad 
= 
M 
n^2 \left( \boldsymbol{r}, t \right) 
\nabla 
\left\{ 
\boldsymbol{v}_s \left( \boldsymbol{r}, t \right) 
\cdot 
\left\lbrack 
\boldsymbol{\Omega}_{\perp} \left( \boldsymbol{r}, t \right) 
\times 
\boldsymbol{r} 
\right\rbrack 
\right\} 
\nonumber\\
&& \qquad 
+ 
M 
n^2 \left( \boldsymbol{r}, t \right) 
\left\lbrack 
\boldsymbol{v}_s \left( \boldsymbol{r}, t \right) 
\cdot 
\nabla 
\Omega_{\perp} \left( \boldsymbol{r}, t \right) 
\right\rbrack 
\boldsymbol{r} 
\times 
\boldsymbol{e}_{\perp} 
\nonumber\\
&& \qquad 
- 
M 
n^2 \left( \boldsymbol{r}, t \right) 
\left\lbrack 
\boldsymbol{r} 
\cdot 
\nabla 
\Omega_{\perp} \left( \boldsymbol{r}, t \right) 
\right\rbrack 
\boldsymbol{v}_s \left( \boldsymbol{r}, t \right) 
\times 
\boldsymbol{e}_{\perp} 
\nonumber\\
&& \qquad 
+ 
M 
n^2 \left( \boldsymbol{r}, t \right) 
\left\{ 
\boldsymbol{e}_{\perp} 
\cdot 
\left\lbrack 
\boldsymbol{r} 
\times 
\boldsymbol{v}_s \left( \boldsymbol{r}, t \right) 
\right\rbrack 
\right\} 
\nabla 
\Omega_{\perp} \left( \boldsymbol{r}, t \right) 
\nonumber\\
&& \qquad 
+ 
M 
n \left( \boldsymbol{r}, t \right) 
\boldsymbol{\Omega}_{\perp} \left( \boldsymbol{r}, t \right) 
\times 
\left\{ 
\left\lbrack 
\boldsymbol{v}_s \left( \boldsymbol{r}, t \right) 
\cdot 
\nabla 
n \left( \boldsymbol{r}, t \right) 
\right\rbrack 
\boldsymbol{r} 
\right\} 
\nonumber\\
&& \qquad 
- 
\frac{1}{2} 
M 
n \left( \boldsymbol{r}, t \right) 
\boldsymbol{\Omega}_{\perp} \left( \boldsymbol{r}, t \right) 
\times 
\left\{ 
\left\lbrack 
\boldsymbol{r} 
\cdot 
\nabla 
n \left( \boldsymbol{r}, t \right) 
\right\rbrack 
\boldsymbol{v}_s \left( \boldsymbol{r}, t \right) 
\right\} 
\nonumber\\
&& \qquad 
- 
\frac{1}{2} 
M 
n \left( \boldsymbol{r}, t \right) 
\boldsymbol{\Omega}_{\perp} \left( \boldsymbol{r}, t \right) 
\times 
\left\{ 
\left\lbrack 
\boldsymbol{r} 
\cdot 
\boldsymbol{v}_s \left( \boldsymbol{r}, t \right) 
\right\rbrack 
\nabla 
n \left( \boldsymbol{r}, t \right) 
\right\} 
\nonumber\\
&& \qquad 
- 
\frac{1}{2} 
M 
n \left( \boldsymbol{r}, t \right) 
\left\{ 
\boldsymbol{\Omega}_{\perp} \left( \boldsymbol{r}, t \right) 
\cdot 
\left\lbrack 
\boldsymbol{r} 
\times 
\nabla 
n \left( \boldsymbol{r}, t \right) 
\right\rbrack 
\right\} 
\boldsymbol{v}_s \left( \boldsymbol{r}, t \right) 
\nonumber\\
&& \qquad 
- 
\frac{1}{2} 
M 
n \left( \boldsymbol{r}, t \right) 
\left\{ 
\boldsymbol{\Omega}_{\perp} \left( \boldsymbol{r}, t \right) 
\cdot 
\left\lbrack 
\boldsymbol{r} 
\times 
\boldsymbol{v}_s \left( \boldsymbol{r}, t \right) 
\right\rbrack 
\right\} 
\nabla 
n \left( \boldsymbol{r}, t \right) 
. 
\nonumber\\
\label{F_G_rot_st1}
\end{eqnarray}

The third to fifth lines in Eq. \eqref{F_G_rot_st1} can be written as

\begin{eqnarray}
&& 
M 
n^2 \left( \boldsymbol{r}, t \right) 
\left\lbrack 
\boldsymbol{v}_s \left( \boldsymbol{r}, t \right) 
\cdot 
\nabla 
\Omega_{\perp} \left( \boldsymbol{r}, t \right) 
\right\rbrack 
\boldsymbol{r} 
\times 
\boldsymbol{e}_{\perp} 
\nonumber\\
&& \; 
- 
M 
n^2 \left( \boldsymbol{r}, t \right) 
\left\lbrack 
\boldsymbol{r} 
\cdot 
\nabla 
\Omega_{\perp} \left( \boldsymbol{r}, t \right) 
\right\rbrack 
\boldsymbol{v}_s \left( \boldsymbol{r}, t \right) 
\times 
\boldsymbol{e}_{\perp} 
\nonumber\\
&& \; 
+ 
M 
n^2 \left( \boldsymbol{r}, t \right) 
\left\{ 
\boldsymbol{e}_{\perp} 
\cdot 
\left\lbrack 
\boldsymbol{r} 
\times 
\boldsymbol{v}_s \left( \boldsymbol{r}, t \right) 
\right\rbrack 
\right\} 
\nabla 
\Omega_{\perp} \left( \boldsymbol{r}, t \right) 
\nonumber\\
&& \quad 
= 
M 
n^2 
\left\lbrack 
\boldsymbol{e}_{\perp} 
\cdot 
\nabla 
\Omega_{\perp} \left( \boldsymbol{r}, t \right) 
\right\rbrack 
\boldsymbol{r} 
\times 
\boldsymbol{v}_s \left( \boldsymbol{r}, t \right) 
, 
\end{eqnarray}
which is zero since $\boldsymbol{e}_{\perp} \cdot \nabla = 0$ in a quasi-2D system.

The sixth to the last lines in Eq. \eqref{F_G_rot_st1} can be written as

\begin{eqnarray}
&& 
M 
n \left( \boldsymbol{r}, t \right) 
\boldsymbol{\Omega}_{\perp} \left( \boldsymbol{r}, t \right) 
\times 
\left\{ 
\left\lbrack 
\boldsymbol{v}_s \left( \boldsymbol{r}, t \right) 
\cdot 
\nabla 
n \left( \boldsymbol{r}, t \right) 
\right\rbrack 
\boldsymbol{r} 
\right\} 
\nonumber\\
&& \; 
- 
\frac{1}{2} 
M 
n \left( \boldsymbol{r}, t \right) 
\boldsymbol{\Omega}_{\perp} \left( \boldsymbol{r}, t \right) 
\times 
\left\{ 
\left\lbrack 
\boldsymbol{r} 
\cdot 
\nabla 
n \left( \boldsymbol{r}, t \right) 
\right\rbrack 
\boldsymbol{v}_s \left( \boldsymbol{r}, t \right) 
\right\} 
\nonumber\\
&& \; 
- 
\frac{1}{2} 
M 
n \left( \boldsymbol{r}, t \right) 
\boldsymbol{\Omega}_{\perp} \left( \boldsymbol{r}, t \right) 
\times 
\left\{ 
\left\lbrack 
\boldsymbol{r} 
\cdot 
\boldsymbol{v}_s \left( \boldsymbol{r}, t \right) 
\right\rbrack 
\nabla 
n \left( \boldsymbol{r}, t \right) 
\right\} 
\nonumber\\
&& \; 
- 
\frac{1}{2} 
M 
n \left( \boldsymbol{r}, t \right) 
\left\{ 
\boldsymbol{\Omega}_{\perp} \left( \boldsymbol{r}, t \right) 
\cdot 
\left\lbrack 
\boldsymbol{r} 
\times 
\nabla 
n \left( \boldsymbol{r}, t \right) 
\right\rbrack 
\right\} 
\boldsymbol{v}_s \left( \boldsymbol{r}, t \right) 
\nonumber\\
&& \; 
- 
\frac{1}{2} 
M 
n \left( \boldsymbol{r}, t \right) 
\left\{ 
\boldsymbol{\Omega}_{\perp} \left( \boldsymbol{r}, t \right) 
\cdot 
\left\lbrack 
\boldsymbol{r} 
\times 
\boldsymbol{v}_s \left( \boldsymbol{r}, t \right) 
\right\rbrack 
\right\} 
\nabla 
n \left( \boldsymbol{r}, t \right) 
\nonumber\\
&& \quad 
= 
\frac{1}{2} 
M 
n \left( \boldsymbol{r}, t \right) 
\Omega_{\perp} \left( \boldsymbol{r}, t \right) 
\left\lbrack 
\boldsymbol{e}_{\perp} 
\cdot 
\nabla 
n \left( \boldsymbol{r}, t \right) 
\right\rbrack 
\boldsymbol{r} 
\times 
\boldsymbol{v}_s \left( \boldsymbol{r}, t \right) 
\nonumber\\
&& \qquad 
+ 
\frac{1}{2} 
M 
n \left( \boldsymbol{r}, t \right) 
\Omega_{\perp} \left( \boldsymbol{r}, t \right) 
\left\lbrack 
\boldsymbol{e}_{\perp} 
\cdot 
\boldsymbol{v}_s \left( \boldsymbol{r}, t \right) 
\right\rbrack 
\boldsymbol{r} 
\times 
\nabla 
n \left( \boldsymbol{r}, t \right) 
, 
\nonumber\\
\end{eqnarray}
which is also zero since $\boldsymbol{v}_s \left( \boldsymbol{r}, t \right) \cdot \boldsymbol{e}_{\perp} = 0$ in a quasi-2D scalar BEC. 
This concludes the derivation of Eq. \eqref{F_G_rot}.

\bibliography{Notes_VE_final}

\end{document}